\documentclass[12pt]{JHEP3}

\usepackage{epsfig,multicol} 

\newcommand{\p}{\partial}

\makeatletter
\def\eqnarray{%
 \stepcounter{equation}%
 \let\@currentlabel=\theequation
 \global\@eqnswtrue
 \global\@eqcnt\z@
 \tabskip\@centering
 \let\\=\@eqncr
 $$\halign to \displaywidth\bgroup\@eqnsel\hskip\@centering
 $\displaystyle\tabskip\z@{##}$&\global\@eqcnt\@ne
 \hfil$\displaystyle{{}##{}}$\hfil
 &\global\@eqcnt\tw@$\displaystyle\tabskip\z@{##}$\hfil
 \tabskip\@centering&\llap{##}\tabskip\z@\cr}
\makeatother

\title{ The Whitham Deformation of 
    the Dijkgraaf-Vafa Theory }

\author{Shogo Aoyama\\
        Department of Physics, Shizuoka University, 
          Ohya 836, Shizuoka, Japan  \\
        E-mail: \email{spsaoya@ipc.shizuoka.ac.jp}}
\author{Takahiro Masuda\\
       Hokkaido Institute of Technology, 
        7-15 Maeda, Teine, Sapporo,  Japan \\
        E-mail: \email{masuda@hit.ac.jp }}

\abstract{We discuss the Whitham deformation of the effective superpotential in the 
Dijkgraaf-Vafa 
(DV) theory. It amounts to discussing the Whitham deformation of an underlying 
(hyper)elliptic curve. Taking the elliptic case for simplicity we derive the Whitham 
equation for the period,  which governs flowings of branch points on the Riemann surface. 
By studying the hodograph solution to the Whitham equation it is shown that the 
effective superpotential in the DV theory
 is realized by  many different meromorphic differentials. Depending on which 
 meromorphic differential to take, the effective superpotential undergoes different 
deformations. This aspect of the DV theory is discussed in detail by taking the $N=1^*$ 
theory. We give a physical interpretation of the deformation parameters.  }

\keywords{Matrix Models, Integrable Hierarchies, Topological Field Theories, Nonperturbative Effects}

\preprint{}

\begin{document}

\section{Introduction}
\setcounter{equation}{0}

One of the long-standing problems in the quantum field theory is to understand 
non-perturbative dynamics of the supersymmetric QCD. The recent works by Dijkgraaf and 
Vafa\cite{DV1, DV2} have given a breakthrough towards to this direction. They found a close 
relation between the $N=1$ supersymmetric QCD and a matrix model. It is now confirmed by 
subsequent works by many people. This Dijkgraaf-Vafa  theory revived  a renewed 
attention\cite{Mir,IT} to the Seiberg-Witten theory (SW) for the  $N=2$ supersymmetric 
QCD\cite{SW}. 
Indeed both theories have  relevance to the Riemann surface and the Whitham hierarchy  as 
commonly underlying features. To be concrete, take the case which is concerned about the 
Riemann surface with genus one. 
 The DV theory gives an effective superpotential of the $N=1$ supersymmetric QCD 
\begin{eqnarray}
 W_{DV} =  \int d^2\theta\ ( N{\partial {\cal F}_0 \over \partial S} 
-2\pi i{\mathop{\tau}^\circ} S),  \label{DVpo}
\end{eqnarray}
which is the perturbative part of the Veneziano-Yankielovicz superpotential. On the other 
hand the SW theory gives the effective action of the $N=2$ supersymmetric QCD
\begin{eqnarray}
 W_{SW} = {1\over 4\pi}{\rm Im}[\int d^4\theta\ {\partial {\cal F}_0(A)\over \partial 
A}\bar A
 + \int d^2\theta\ {1\over 2}\tau_0(A) W_\alpha W^\alpha ].  \label{SWpo}
\end{eqnarray}
In (\ref{DVpo}) and  (\ref{SWpo}) ${\cal F}_0$ is called the free energy and  characterized 
by a certain differential on an elliptic curve. 
 In (\ref{SWpo}) $\tau_0(A)$  is the period of the curve. 

\vspace{0.5cm}

Since the SW theory was born,  the Whitham hierarchy attracted attention as a underlying  
integrable structure  in the SW theory\cite{Mar,IM,SW2}. But dynamical aspects of the 
hierarchy was not properly studied. Dynamical flowings of branch points of the Riemann 
surface and the consequent creation or annihilation of branch cuts were the original 
concerns to  study the   Whitham hierarchy\cite{Whi}.  There people aimed regulation (or 
modulation) of  singular behaviors of finite-gap solutions for the KdV system. We think it 
important to shed  more light on this aspect of the DV and SW theories in the revived era 
of the interest.   In this paper we study the Whitham deformation of the elliptic curve as 
a flow of the branch points on the Riemann surface. 
 We will show that the flow is 
governed by the Whitham equation for the period $\tau$ of the curve (\ref{whitt}), {\it 
i.e.},
\begin{eqnarray}
 {\partial \tau (a,\vec T) \over \partial T_M} = s_M(\tau(a,\vec T)) {\partial \tau(a,\vec 
T) \over \partial a},  \quad\quad  \tau (a,\vec 0) \equiv \tau_0(a).    \label{*}
\end{eqnarray}
Here  $\vec T = (T_1,T_2,\cdots)$  and $a$ is either $S$ for the DV theory or $A$ for the 
SW theory. $s_1(\tau(a,\vec T)), s_2(\tau(a,\vec T)),\cdots $,  are the characteristic 
speeds. (For more explanations see the discussion thereabout.)
  It is noteworthy that the Whitham equation  looks like the renormalization group equation 
for the running coupling 
$$
{\partial \bar g(t,g) \over \partial t} = \beta(g){\partial \bar g(t,g) \over \partial g}.
$$

\vspace{0.5cm}

In the recent papers\cite{Hollow} they developed interesting arguments on an underlying 
integrable structure of  the DV theory. Namely  they discussed that the equilibria of the 
superpotential of the DV theory, (\ref{DVpo}), correspond to the stable flow points of some 
integrable systems at which the relevant  (hyper)elliptic curve degenerates. 
 But the flow is not the one given by the Whitham deformation (\ref{*}), which is of 
interest in this paper.

\vspace{0.5cm}

It is known\cite{KG} that the equation of the type (\ref{*}) admits a hodograph solution as 
a special solution and it is characterized by a free energy ${\cal F}$. 
We will find that the initial condition $\tau_0(a)$ of a hodograph solution   
is imposed by specifying the inverse function $a(\tau_0) (\equiv \tau_0^{-1}(a))$ as 
(\ref{4.5}), {\it i.e.}, 
\begin{eqnarray}
a(\tau(a,\vec 0)) = \sum_{M=1}^{M_0}\Lambda_M s_M(\tau(a,\vec 0)).   \label{in}
\end{eqnarray}
 It will be shown that (\ref{in}) is given as a period integral of a certain differential, 
which is  nothing but the DV or SW differential. Then 
the  free energy ${\cal F}_0$ in (\ref{DVpo})  and (\ref{SWpo}) is an initial value of the 
%\keywords{mmo, ihi, tpf, nef}
free energy ${\cal F}$ which characterizes the hodograph solution
\begin{eqnarray}
{\cal F}_0(a) = {\cal F}|_{\vec T= 0}. \label{free}
\end{eqnarray}

\vspace{0.5cm}%\keywords{mmo, ihi, tpf, nef}

 Choosing a set of non-vanishing $\Lambda_M$ in (\ref{in}) 
uniquely specifies the initial condition $\tau_0 (a)$ and therefore the free energy ${\cal 
F}_0(a)$. Here the characteristic speeds $s_M(\tau(a,\vec 0)), M=1,2,\cdots$, were  kept 
fixed. 
Now  we reverse the argument. Namely  we  fix the initial condition   
$\tau_0 (a)$, that is, the {\it l.h.s.} of (\ref{in}), and  specify the characteristic 
speeds  so as to satisfy (\ref{in}) for the chosen set of 
$(\Lambda_1,\Lambda_2,\cdots,\Lambda_{M_{0}})$. We show that it can be done by changing  
parameterization of 
the elliptic curve. 
Then the fixed initial condition $\tau_0 (a)$ is  given by different DV or SW 
differentials. The hodograph solution $\tau(a,\vec T)$ is characterized by different free 
energies ${\cal F}$. But the key point is that its  initial value ${\cal F}_0(a)$ remains 
fixed as $\tau_0 (a) = {\p^2 {\cal F}_0 \over \p a^2}$. In other words a 
fixed  ${\cal F}_0(a)$  undergoes different Whitham deformations depending on the choice of  
$(\Lambda_1,\Lambda_2,\cdots,\Lambda_{M_{0}})$. 
To discuss concretely this aspect  of the DV theory, we will take the $N=1^*$ 
theory\cite{DV2,DV3} as an example. We will explicitly give the  superpotential of the 
theory  
by various DV differentials, and show  the possibility of different Whitham deformations.

\vspace{0.5cm}

The paper is organized as follows. In Section 2 we define quantities $\tau_{MN}$ as 
extensions of the period $\tau (\equiv \tau_{00})$ of the elliptic curve. 
 They are important constituents in the paper.
 In Section 3 we discuss on a ``gauge" freedom in parameterizing the elliptic curve. Fixing 
this freedom we find an equation for the  curve, which plays a fundamental role 
in the paper. 
In Section 4 we  think of deforming the  curve. We introduce the Whitham hierarchy 
to the deformation. It is  shown to amount to assuming  the Whitham equation (\ref{*}). The 
Whitham hierarchy is mimic to the dispersionless KP hierarchy in the topological field 
theory\cite{LG,Eguchi,SK}. Therefore it may be formulated by means of analogous quantities 
with the 2-point 
functions of the topological field theory. We find them  by modifying $\tau_{MN}$, 
discussed in Section 2. 
 In Section 5  we study the hodograph solution to the Whitham equation according to 
\cite{KG,SK}. We then interpret the solution in terms of the topological field theory, {\it 
i.e.}, by employing the terminology  like
 the free energy, the {\it small} or {\it large phase space} {\it etc}. In Section 6 the 
dual version of the Whitham equation is discussed. The initial condition for the hodograph 
solution is characterized by the DV or SW differential. In the basis of  these arguments  
we  discuss various Whitham deformations of the $N=1^*$ theory in Section 7.  
In Section 8 we  give a matrix-model interpretation of   those Whitham deformations. 
 Appendix A is 
devoted to give a short summary on the elliptic curve and some useful formulae for the 
period integral. In Appendix B we discuss a systematic method to evaluate the 2-point 
functions $\tau_{MN}$. In Appendix C we give some calculations to  check the consistency 
of the formalism for the Whitham hierarchy, developed in this paper.

\vspace{1cm}

\section{The elliptic curve}
\setcounter{equation}{0}

Throughout the paper we consider an elliptic curve  defined by
\begin{eqnarray}
 y^2 = 4(x-u)(x-v)(x-w).  \label{2.1}
\end{eqnarray}
It may be also given by 
$$
y^2 = 4(x-\lambda_1)(x-\lambda_2)(x-\lambda_3)(x-\lambda_4).
$$
But this reduces to  (\ref{2.1}) by the change
$$
{1\over x-\lambda_4} \longrightarrow x, \quad\quad\quad 
{y\over \sqrt {(\lambda_4-\lambda_1)(\lambda_4-\lambda_2)(\lambda_4-\lambda_3)}} 
\longrightarrow {y\over x^2}.
$$
On the curve (\ref{2.1}) there exists one holomorphic differential $d\omega_0$ of the 
form 
\begin{eqnarray}
d\omega_0 = {1\over g_0}{dx\over y}.  \label{2.2}
\end{eqnarray}
By the normalization 
\begin{eqnarray}
\oint_A d\omega_0 = 1,   \label{2.3}
\end{eqnarray}
$g_0$ is fixed to be 
\begin{eqnarray}
g_0 = \oint_A {dx\over y}.    \label{2.4}
\end{eqnarray}
The period on the curve is given by 
\begin{eqnarray}
\tau &=& \oint_B d\omega_0 = {g_{D0}\over g_0},   \label{2.6}
\end{eqnarray}
with
\begin{eqnarray}
g_{D0} = \oint_B {dx\over y}.    \nonumber
\end{eqnarray}
On the curve  there exists also a set of meromorphic differentials $d\Omega_M, 
M=1,2,\cdots,$ of the form 
\begin{eqnarray}
d\Omega_M &=& [x^M + \gamma_{M-1}x^{M-1}+\cdots +\gamma_1 x + \gamma_0]{dx\over y} 
\label{2.7}  \\
& \equiv & Q_M(x){dx\over y}. \nonumber
\end{eqnarray}
The coefficients $\gamma_{M-1},\cdots ,\gamma_0$ of the polynomial $Q_M$ is uniquely 
determined by requiring $d\Omega_M$ to satisfy  
\begin{eqnarray}
\oint_A d\Omega_M = 0,      \label{2.8}
\end{eqnarray}
and to have poles at $x ={1\over \xi^2}= \infty$ such that 
\begin{eqnarray}
d\Omega_M = -\xi^{-2M}d\xi +  holomorphic.  \label{2.9}
\end{eqnarray}
It is done by the recursive formula (\ref{A.4}) and (\ref{gamma00}). 

\vspace{0.5cm}

 Let us  introduce the following quantities as an extension of the period (\ref{2.6}):
\begin{eqnarray}
\tau_{0M} & \equiv & \tau_{M0}      \nonumber  \\
          & = & \oint_B d\Omega_M, \quad\quad\quad M = 1,2,\cdots. \label{2.12}
\end{eqnarray}
It becomes 
\begin{eqnarray}
 \tau_{0M}= -2\pi i \ {\rm res}_{\xi=0}[{\scriptstyle{\xi^{-2M+1}}\over \scriptstyle 
{2M-1}}d\omega_0], \label{2.13}
\end{eqnarray}
by the Riemann bilinear relation. By the analogy  we also introduce the quantities
\begin{eqnarray}
\tau_{NM} = -2\pi i\ {\rm res}_{\xi=0}[{\scriptstyle{\xi^{-2M+1}}\over 
\scriptstyle{2M-1}}d\Omega_N],\quad\quad  M,N = 1,2,\cdots.
 \label{2.14}
\end{eqnarray}
Using the Riemann bilinear relation again we have
\begin{eqnarray}
\tau_{NM} = \tau_{MN}.  \label{tauNM}
\end{eqnarray}
These quantities will be important constituents in our discussions later.

\vspace{1cm}

\section{Gauge fixing and the fundamental equation}
\setcounter{equation}{0}

Let us assume that the branch points of  (\ref{2.1}) are parameterized by one parameter 
alone, namely  the period as
$ u(\tau), v(\tau), w(\tau)$. In general an elliptic curve is determined  by $g_0$ and 
$g_{D0}$. The assumption implies that they are  functions of $\tau$. To realize the 
assumption 
it suffices to set $g_0$ to be a particular function of $\tau$, because they are related by 
(\ref{2.6}). We call it ``gauge fixing", which is an abusive terminology. For instance, 
with the ``gauge" $g_0 = 2\pi$ the 
branch points of the curve are expressed as
\begin{eqnarray}
 u(\tau) &=& {c\over 3} + {1\over 12}[\theta_3(\tau)^4 + \theta_0(\tau)^4 ], \nonumber \\
v(\tau) &=& {c\over 3}+ {1\over 12}[\theta_2(\tau)^4 - \theta_0(\tau)^4 ], \label{branch1} 
\\
w(\tau) &=&  {c\over 3}- {1\over 12}[\theta_2(\tau)^4 + \theta_3(\tau)^4 ], \nonumber 
\end{eqnarray}
by using the Weierstrass standard form (\ref{C.1}) and (\ref{C.2}).
 Here $c$ is given by
\begin{eqnarray}
c=u(\tau) +v(\tau) + w(\tau),  \label{defc}
\end{eqnarray}
which is still to be fixed arbitrarily. We can also take the ``gauge" which sets one of the 
branch points, for instance, $w(\tau)=1$. Then we have 
\begin{eqnarray}
 u(\tau) &=& {c\over 3} + {1\over 3}({\pi\over g_0})^2[\theta_3(\tau)^4 + 
\theta_0(\tau)^4 ], \nonumber \\
v(\tau) &=& {c\over 3}+ {1\over 3}({\pi\over g_0})^2[\theta_2(\tau)^4 - 
\theta_0(\tau)^4 ], \label{branch2}
\end{eqnarray}
where $g_0$ is given by
$$
1 = {c\over 3}- {1\over 12}({\pi\over g_0})^2[\theta_2(\tau)^4 + 
\theta_3(\tau)^4 ].
$$
With the former ``gauge" the three branch points $u(\tau), v(\tau), w(\tau)$ move depending 
$\tau$. Instead  with the latter one  only two of them move. 

\vspace{0.5cm}

When the branch points $ u(\tau), v(\tau), w(\tau)$ are given as such, we can show  the 
fundamental equation throughout the paper:
\begin{eqnarray}
 {\partial \over \partial \tau}d\Omega_M(x) =
 s_M(\tau) {\partial \over \partial \tau}d\omega_0 + d[\Delta_M(x,\tau)].    
  \quad\quad  M=1,2,\cdots.
\label{whit}
\end{eqnarray}
Here $s_M(\tau)$ in the first term is  the characteristic speed given by 
\begin{eqnarray}
s_M(\tau) = g_0 {Q_M(u)u'(v-w) + Q_M(v)v'(w-u) + Q(w)_Mw'(u-v)  \over u'(v-w) +v'(w-u) + 
w'(u-v) },
   \label{speed} 
\end{eqnarray}
and the second term is an exact form with 
\begin{eqnarray}
\Delta_M(x,\tau) = &-&{Q_M(u)u'(v'w-w'v) + Q_M(v)v'(w'u-u'w) + Q_M(w)w'(u'v-v'u) \over
  u'(v-w) +v'(w-u) + w'(u-v) } {1\over y}   \nonumber \\
&+& {Q_M(u)u'(v'-w') + Q_M(v)v'(w'-u') + Q_M(w)w'(u'-v') \over  u'(v-w) +v'(w-u) + w'(u-v) 
} {x\over y}.   \label{exact}
\end{eqnarray}
To prove (\ref{whit}), first of all note that (\ref{2.9}) implies that 
${\partial \over \partial \tau}d\Omega_M(x)$ is holomorphic. Therefore  we have 
\begin{eqnarray}
 &\quad& {\partial \over \partial \tau}d\Omega_M(x)  \nonumber \\
&\quad&  \hspace{1cm} = {\partial Q_M(x)\over \partial \tau}{dx \over y} +Q_M(x) {\partial 
\over \partial \tau}{dx \over y} \nonumber \\
&\quad&  \hspace{1cm} = [A_M (u,v,w)    + {1\over 2}({u'Q_M(u) \over x-u} + {v'Q_M(v)\over 
x-v} + {w'Q_M(w)\over x-w})]{dx\over y}, \label{diform}
\end{eqnarray}
with some functions $A_M (u,v,w) $. On the other hand we also note that
\begin{eqnarray}
{1\over x- v}{dx\over y} &=& {1\over w-v}(1-{w-u\over x-u}){dx\over y} + 
d[{2\over v-w}({w\over y}-{x\over y})], \nonumber \\
{1\over x-w}{dx\over y} &=& {1\over v-w}(1-{v-u\over x-u}){dx\over y} 
+ d[{2\over w-v}({v\over y}-{x\over y})],  \label{f}
\end{eqnarray}
by calculating $d({1\over y})$ and  $d({x\over y})$. Plugging these into 
 the equation, obtained by calculating ${\p \over \p \tau}{dx \over y}$ similarly to 
(\ref{diform}), we find  
\begin{eqnarray}
{\partial \over \partial \tau}{dx\over y} 
 = \alpha(u,v,w){dx\over y} + \beta(u,v,w){1\over x-u}{dx\over y}
  + d[\delta(u,v,w)],   \label{difholo}
\end{eqnarray}
with
\begin{eqnarray}
\alpha(u,v,w) &=& -{1\over 2} {v'-w'\over v-w}, \nonumber \\
\beta(u,v,w) &=& {1\over 2}(u'-v'{w-u\over w-v}-w'{v-u\over v-w}), \nonumber \\
\delta(u,v,w)&=& {v'w-w'v \over v-w}{1\over y} - {v'-w' \over v-w}{x\over y}. 
\nonumber
\end{eqnarray}
By (\ref{difholo}) as well as those obtained by replacing $u,v$ and $w$ cyclically, 
(\ref{diform}) becomes 
\begin{eqnarray}
{\partial \over \partial \tau}d\Omega_M(x) = \tilde A_M(u,v,w){dx\over y} 
+ B_M(u,v,w){\p \over \p \tau}{dx\over y} + d[C_M(u,v,w)],  \label{deriva}
\end{eqnarray}
with 
\begin{eqnarray}
\tilde A_M(u,v,w) &=& A_M(u,v,w) -{1\over 2}[u'Q_M(u){\alpha(u,v,w)\over \beta(u,v,w)} + 
cyclic],  \nonumber\\
B_M(u,v,w) &=& {1\over 2}[u'Q_M(u) { 1 \over \beta(u,v,w)} + cyclic], \nonumber \\
C_M(u,v,w) &=& -{1\over 2}[u'Q_M(u){ \delta (u,v,w) \over \beta(u,v,w)} + cyclic]. 
\nonumber
\end{eqnarray}
Integrating (\ref{deriva}) along a $A$-cycle  yields 
$$
\tilde A_M(u,v,w) = - B_M(u,v,w) {g_0'\over g_0}, \quad\quad M = 1,2,\cdots,
$$
owing to (\ref{2.8}). 
Plug this into (\ref{deriva}). Then we obtain the fundamental equation (\ref{whit}).

\vspace{1cm}

\section{The Whitham deformation}
\setcounter{equation}{0}

We consider a deformation of a ``gauge"-fixed curve 
\begin{eqnarray}
y^2 = 4(x-u(\tau))(x-v(\tau))(x-w(\tau)),     \label{gcurve}
\end{eqnarray}
through  that of  $\tau$:
\begin{eqnarray}
\tau \longrightarrow \tau(a,T_1,T_2,\cdots\cdots) \equiv \tau(a,\vec T),  \nonumber
\end{eqnarray}
with flow parameters $a$ and $T_M,\ M=1,2,\cdots$. The Whitham deformation is defined  by
\begin{eqnarray}
{\p \over \p T_M}d\omega_0 &=& {\p \over \p a}d\tilde \Omega_M,  \label{3.1} \\
{\partial \over \partial T_M}d\tilde\Omega_N &=& {\partial \over \partial 
T_N}d\tilde\Omega_M,   \label{3.1'}
\end{eqnarray}
in which $d\tilde \Omega_M$ is a  modified meromorphic differential of $d\Omega_M$, 
(\ref{2.7}), as
\begin{eqnarray}
d\tilde \Omega_M = d\Omega_M - d(\int^\tau d\tau\ \Delta_M ), \label{Omega'}
\end{eqnarray}
with  $\Delta_M$ given by (\ref{exact}). The compatibility of the deformation can be seen 
by showing that both (\ref{3.1}) and(\ref{3.1'})  are equivalent to  the  equation
\begin{eqnarray}
{\p  \over \p T_M}\tau(a,\vec T) = s_M(\tau){\p  \over \p a}\tau(a,\vec T) .   
\label{whitt}
 \label{3.2}
\end{eqnarray}
It is called the Whitham equation.  We calculate the {\it r.h.s.} of (\ref{3.1})  with 
(\ref{whit}): 
$$
{\p \over \p a}d\tilde\Omega_M = {\p \tau \over \p a}{\p \over \p \tau}d\tilde\Omega_M = 
{\p \tau \over \p a}s_M(\tau) {\p \over \p \tau}d\omega_0,
$$
which is equal to the {\it l.h.s.} of (\ref{3.1})  due to the Whitham equation (\ref{3.2}). 
Similarly we calculate the {\it l.h.s.} of (\ref{3.1'}) by (\ref{whit}) and (\ref{3.2}):
$$
{\partial \over \partial T_M}d\tilde\Omega_N = {\partial \tau \over \partial T_M}{\partial 
\over \partial \tau}d\tilde\Omega_N = s_M(\tau)s_N(\tau){\partial \over \partial 
a}d\omega_0.
$$
This implies (\ref{3.1'}). Thus all of the flows defined by (\ref{3.1}) and(\ref{3.1'}) are 
compatible and they are equivalent to the the single equation (\ref{whitt}).

\vspace{0.5cm}

By (\ref{2.6}), (\ref{2.12})$\sim$(\ref{2.14}) and (\ref{Omega'}),  we write (\ref{3.1}) 
and(\ref{3.1'}) as 
\begin{eqnarray}
{\partial \tau \over \partial T_M} &=& {\partial \tau_{0M}\over \partial a},
 \label{3.1''}  \\
{\p \tau_{0K} \over \p T_M} &=& {\p \tilde \tau_{MK} \over \p a} \label{3.2'} \\
{\p \tilde\tau_{NK}\over \p T_M} &=& {\p \tilde\tau_{MK}\over \p T_N},
  \label{3.4}
\end{eqnarray}
with  
\begin{eqnarray}
\tilde \tau_{NM}&=& -2\pi i \ {\rm res}_{\xi=0}[{\scriptstyle{\xi^{-2M+1}}\over 
\scriptstyle{2M-1}}d\tilde \Omega_N] \nonumber \\
      &=& \tau_{NM}+2\pi i\ {\rm res}_{\xi=0}[\xi^{-2M}d\xi \int^\tau d\tau\ 
\Delta_N(x,\tau)],   
\label{tildetau}
\end{eqnarray}
for $M,N=1,2,\cdots$. By the Riemann bilinear relation we have also for $\tilde \tau_{NM}$ 
$$
\tilde \tau_{NM} = \tilde \tau_{MN}.
$$
(\ref{3.1''})$\sim$(\ref{3.4}) imply that $\tilde\tau_{MN}$ together with $\tau(\equiv 
\tau_{00})$ and $\tau_{0M}(\equiv \tilde\tau_{0M})$ are integrable as
\begin{eqnarray}
\tilde\tau_{AB} = {\p^2 {\cal F}\over \p T_A \p T_B}, \quad\quad
 A,B = 0,1,2,\cdots,       \label{3.5}
\end{eqnarray}
with a function ${\cal F}$ of $T_0(=a)$ and $\vec T$.
 This Whitham hierarchy has exactly the same integrable structure as the dispersionless KP 
hierarchy for the topological Landau-Ginzburg theory\cite{LG,Eguchi}. Employing the 
terminology in 
the latter theory we call ${\cal F}$ the free energy and $\tilde\tau_{AB}$ the $2$-point 
function. 

\vspace{0.5cm}

Thus the Whitham hierarchy is mimic to the dispersionless KP hierarchy. 
In the rest of this section and the next section we proceed the arguments by taking the 
close analogy with the topological field theory.

\vspace{1cm}

\section{The hodograph solution}
\setcounter{equation}{0}
 
In this section we solve the Whitham equation (\ref{whitt}) in order to see the   flow. 
 It is known in  \cite{KG,SK} that for an equation of the sort (\ref{whitt}) we have a 
hodograph solution such as given by 
\begin{eqnarray}
\tau(a,\vec T) \equiv \hat \tau (\hat a),      \label{4.1}
\end{eqnarray}
where 
\begin{eqnarray}
\hat \tau(a) & \equiv & \tau(a,\vec 0),    \label{4.2} \\
\hat a & =& a + \sum_{M=1}^{\infty} T_M s_M(\tau(a,\vec T)).  \label{4.3}
\end{eqnarray}
To show this write (\ref{whitt}) in the form 
\begin{eqnarray}
[{\p \over \p T_M} - s_M(\tau){\p \over \p a }]\tau = 0.
\end{eqnarray}
This implies that $\tau$ is constant along the characteristic 
\begin{eqnarray}
{d\ T_M \over -1} = {d a \over s_M(\tau)},    \label{4.4}   
\end{eqnarray}
which is a straight line. 
Therefore (\ref{4.1}) with (\ref{4.2}) and (\ref{4.3}) is a solution of the Whitham 
equation. In (\ref{4.2}) $\tau(a,\vec 0)$ is an arbitrary function. Impose the relation
\begin{eqnarray}
a = \sum_{M=1}^{M_0}\Lambda_M s_M(\tau(a,\vec 0)),   \label{4.5}
\end{eqnarray}
with a finite number of constants $\Lambda_M,\ M=1,2,\cdots,M_0$. Then $\tau(a,\vec 0)$ is 
given by inverting (\ref{4.5}). The hodograph solution (\ref{4.1}) is an implicit solution. 
It still has dependence on the solution $\tau(a,\vec T)$ itself through $\hat a$. An 
explicit form of the solution  is obtained in a formal series of $T_M$ :
\begin{eqnarray}
\tau(a,\vec T) &\equiv& \hat \tau(\hat a)  \nonumber \\
 &=& \tau(a,\vec 0) + {\p \tau(a,\vec 0) \over \p a}
\sum_{M=1}^\infty T_M s_M(\hat \tau)       \label{4.6}  \\
&\quad&  \quad\quad\quad  + {1\over 2!} {\p^2 \tau(a,\vec 0)\over \p a^2}
\sum_{M,N=1}^\infty T_MT_N s_M(\hat \tau)s_N(\hat \tau) + \cdots,    \nonumber
\end{eqnarray}
where $s_M(\hat \tau)$ should be also made explicit by iterating the expansion
\begin{eqnarray}
s_M(\hat \tau) &=& s_M(\tau(a,0)) + {\p s_M(\tau(a,0))\over \p a}\sum_{M=1}^\infty T_M 
s_M(\hat \tau)      \nonumber  \\
&\quad&  \quad\quad\quad\quad\quad + {1\over 2!}{\p^2 s_M(\tau(a,0))\over \p a^2}  \sum_{M,
N=1}^\infty T_MT_N s_M(\hat \tau)s_N(\hat \tau)
 + \cdots          \nonumber
\end{eqnarray}
 Thus we see that the initial function  $\tau(a,\vec 0)$ is a generator of the hodograph 
solution (\ref{4.1}).

\vspace{0.5cm}

With the replacement $a$ by $\hat a$ (\ref{4.5}) becomes  
\begin{eqnarray}
\hat a = \sum_{M=1}^{M_0}\Lambda_M s_M(\hat \tau(\hat a)).   \label{4.7} 
\end{eqnarray}
Combining (\ref{4.3}) and (\ref{4.7}) we have 
\begin{eqnarray}
a + \sum_{M=1}^{M_0}\tilde T_M s_M(\tau(a,\vec T)) = 0,   \label{4.8}
\end{eqnarray}
with $\tilde T_M = T_M - \Lambda_M$. This is a constraint satisfied by the hodograph 
solution, which was called  the string equation in the topological field theory\cite{LG}. 
Multiplying (\ref{4.8}) by ${\p \tau \over \p a}$ and using the Whitham equation 
(\ref{whitt}) gives 
\begin{eqnarray}
[a{\p \over \p a } + \sum_{M=1}^{M_0}\tilde T_M {\p \over \p T_M} ]\tau = 0.
\label{4.9}
\end{eqnarray}  
From this 
\begin{eqnarray}
\sum_{A=0}^\infty \tilde T_A{\p \tilde\tau_{BC}\over \p T_A}  = 0, 
 \label{4.10}
\end{eqnarray}
with $\tilde T_0 = a$, since $\tilde \tau_{BC}$ 
are functions of $\tau$. 
By means of (\ref{4.10}) we can easily show that the free energy in (\ref{3.5})  takes the 
form
\begin{eqnarray}
{\cal F} = \sum_{A,B=0}^\infty {1\over  2}\tilde T_A \tilde T_B \tilde\tau_{AB}(\tau(a,\vec 
T)),
  \label{4.11}
\end{eqnarray}
{\it modulo}  linear terms in $T_A$. Putting the hodograph solution (\ref{4.6}) in 
(\ref{4.11}) yields the free energy in a formal series of $T_A$. When $T_M = 0$,\  
$^\forall M$, it becomes 
\begin{eqnarray}
{\cal F}_0(a) = {1\over 2}a^2 \tau((a,\vec 0)) - \sum_{M=1}^{M_0} a\Lambda_M 
\tau_{0M}(\tau(a,\vec 0)) + {1\over 2}\sum_{M,N=1}^{M_0} \Lambda_M\Lambda_N \tilde\tau_{MN}
(\tau(a,\vec 0)).  \nonumber\\
    \label{4.12}
\end{eqnarray}

\vspace{0.5cm}

We finally note that the free energy (\ref{4.11}) satisfies 
\begin{eqnarray}
2{\cal F} = \sum_{A=0}^\infty \tilde T_A{\p {\cal F}\over \p T_A} ,  \label{4.13}
\end{eqnarray}
and
\begin{eqnarray}
{\p {\cal F}\over \p T_B} = \sum_{A=0}^\infty \tilde T_A \tilde\tau_{AB},  \label{4.14}
\end{eqnarray}
by (\ref{4.10}).

\vspace{0.5cm}

In \cite{SK} the hodograph solution  for the dispersionless KP hierarchy was 
interpreted in terms of the topological field theory. The same interpretation is applicable 
also for the case of the Whitham hierarchy. Namely 
the  solution $\tau(a,\vec T) $ and the corresponding free energy ${\cal F}$ are regarded 
 as flowing in the {\it large phase space} of the scaling parameters $T_M$. On 
the other hand the initial values  $\tau_0 (a)$ and ${\cal F}_0(a)$ are regarded as  
flowing  in the {\it small phase space} where the flow parameters $\tilde T_M (= 
T_M -\Lambda_M) $
 are fixed to be $-\Lambda_M$.  Such a {\it small phase space} may be associated with a 
Higgs 
vacuum in the quantum field theory, because $\tau_0 (a)$ depends on the set of 
non-vanishing $\Lambda_M$ and 
the  hodograph solution is perturbatively  constructed from $\tau_0 (a)$  as 
(\ref{4.6}). This aspect will be discussed in Section 8 by using a matrix model. 

\vspace{1cm}

\section{The dual Whitham equation}
\setcounter{equation}{0}

In this section we discuss a dual version of the Whitham equation (\ref{whitt}). Let us 
take a Legendre transform of the flow parameter $a$ to $a_D$ by
\begin{eqnarray}
a_D = {\p {\cal F}\over \p a} .    \label{5.1}
\end{eqnarray}
We then consider $\tau$ as a function of $a_D$ and $\vec T$.
 By differentiating $\tau$ as 
\begin{eqnarray}
{\p \tau \over \p T_M} = {\p \tau \over \p a_D} \Bigl({\p a_D \over \p T_M}\Bigr)_{a={\rm 
const}}
 + \quad \Bigl({\p \tau \over \p T_M}\Bigr)_{a_D={\rm const}},  \nonumber
\end{eqnarray}
the Whitham equation (\ref{whitt}) becomes 
\begin{eqnarray}
\Bigl({\p \tau \over \p T_M}\Bigr)_{a_D={\rm const}} = s_{DM}(\tau){\p \tau \over \p a_D},  
\label{5.2}
\end{eqnarray}
with
\begin{eqnarray}
s_{DM}(\tau) = \tau s_M(\tau) - \tau_{0M}.   \label{5.3}
\end{eqnarray}
Here use is made of (\ref{3.5})  and 
\begin{eqnarray}
{{\p a_D \over \p \tau} \over {\p a \over \p \tau}} = {\p a_D \over \p a} = \tau.   
\label{5.4}
\end{eqnarray}
The last equality  follows from (\ref{5.1}) and (\ref{3.5}). $s_{DM}(\tau)$ is a  dual 
version of the characteristic speed (\ref{speed}). With this form of $s_{DM}(\tau)$, 
(\ref{5.2}) may be called the dual Whitham equation.

\vspace{0.5cm}

The Legendre transform (\ref{5.1}) also induces the dual version of other equations. For 
instance, from (\ref{4.8}) and (\ref{5.3}) we have
\begin{eqnarray}
a\tau + \sum_{M=1}^\infty \tilde T_M ( s_{DM}(\tau) + \tau_{0M} ) = 0, \nonumber
\end{eqnarray}
which becomes
\begin{eqnarray}
a_D + \sum_{M=1}^\infty \tilde T_M s_{DM}(\tau)  = 0,  \label{5.6}
\end{eqnarray}
by (\ref{4.14}). In the case when $T_M = 0$, \  $^\forall M$, (\ref{5.6}) is
reduced to
\begin{eqnarray}
a_D = \sum_{M=1}^{M_0} \Lambda_M s_{DM}(\tau(a,\vec 0)),  \label{5.7}
\end{eqnarray}
which is dual to the relation (\ref{4.5}). They  are paired as 
\begin{eqnarray}
a &=& \oint_A dS,    \label{5.8} \\
a_D &=& \oint_B dS,   \label{5.9} 
\end{eqnarray}
with the differential $dS$ 
\begin{eqnarray}
dS = \sum_{M=1}^{M_0} \Lambda_M [ s_M(\tau)d\omega_0 -d \tilde\Omega_M ],
 \label{5.10}
\end{eqnarray}
owing to (\ref{2.2}), (\ref{2.13}) and (\ref{Omega'}). By (\ref{5.3}) 
we remark that 
\begin{eqnarray}
a_D = \tau a -  \sum_{M=1}^{M_0} \Lambda_M \tau_{0M},  \label{5.10'}
\end{eqnarray}
in the case when  $T_M = 0$, \  $^\forall M$.  
In the next section we will show that the  Seiberg-Witten and Dijkgraaf-Vafa differentials 
are respectively obtained  as special cases of the differential (\ref{5.10}).

\vspace{0.5cm}

The Legendre transform of the free energy ${\cal F}$  is given by 
\begin{eqnarray}
{\cal F}_D = a a_D - {\cal F}.   \nonumber
\end{eqnarray}
We can easily show that 
\begin{eqnarray}
{\p^2 {\cal F}\over \p a^2}  &=& \tau,   \label{5.11} \\ 
{\p {\cal F}_D\over \p a_D } &=& a,  \quad\quad\quad
{\p^2 {\cal F}_D\over \p a_D^2}   = {1\over \tau}, \quad\quad etc.
\nonumber 
\end{eqnarray}
The same formulae hold for ${\cal F}_0$ and ${\cal F}_{0D}$. 
Particularly the following formulae are useful later:
 \begin{eqnarray}
{\partial  {\cal F}_0 \over \partial a } &=& \oint_B dS,  \label{formula1} \\
{\partial  {\cal F}_0 \over \partial (-\Lambda_M) } &=& -2\pi i\ {\rm res}_{\xi=0}
[ {\scriptstyle{\xi^{-2M+1}}\over \scriptstyle{2M-1}}dS],  \label{formula2}
\end{eqnarray}
which can be shown from (\ref{4.14}) by means of (\ref{2.12})$\sim$(\ref{tauNM}) and 
(\ref{5.10}).

\vspace{0.5cm}

 We see that (\ref{5.8}) with the differential $dS$ (\ref{5.10}) is another expression of 
the initial constraint (\ref{4.5}), which was inverted to give 
 the initial condition $\tau_0(a)$ for the hodograph solution.
  Choosing a different set of non-vanishing $\Lambda_M$ specifies the differential $dS$. 
Correspondingly the free energy ${\cal F}_0$ is fixed according to 
(\ref{4.12}). When the flow parameters $T_M, M=1,2,\cdots,$ are turned on, ${\cal F}_0$ 
flows in the {\it large phase space} as given by  ${\cal F}$ with (\ref{4.6}). 

\vspace{1cm}

\section{Applications}
\setcounter{equation}{0}

\subsection{$N=2$ effective  Yang-Mills theory with $SU(2)$}

\noindent
The relevant curve takes the form  
\begin{eqnarray}
y^2 = 4(x^2-1)(x-u). \nonumber
\end{eqnarray}
We put it in the Weierstrass standard form (\ref{C.1}) with
\begin{eqnarray}
e_1 = {2\over 3}u,\quad\quad\quad 
e_2 = 1 - {u\over 3}, \quad\quad\quad 
e_3 = -1 - {u\over 3}.  \nonumber
\end{eqnarray}
A simple manipulation of (\ref{C.2}) gives the relations
\begin{eqnarray}
u &=& -1 + 2\left[{\theta_3(\tau) \over \theta_2(\tau)}\right]^4, \label{C.4'}\\\nonumber 
\\
g_0 &=& -2\omega_1 = {\pi \over \sqrt 2}\theta_2(\tau)^2. \label{C.4''} 
\end{eqnarray}
We consider the hodograph solution of the Whitham equation (\ref{whitt}), imposing  the 
initial condition (\ref{4.5}) with $\Lambda_M = 0$ for $M \ge 2$, {\it i.e.}, 
\begin{eqnarray}
a = \Lambda_1 s_1(\tau). \label{a'}
\end{eqnarray}
Here $s_1(\tau)$, given by (\ref{speed}), can be calculated as
\begin{eqnarray}
s_1(\tau) = g_0Q_1(u) = g_0 (u + \gamma_0), \nonumber
\end{eqnarray}
with
$$
\gamma_0 = - {1\over g_0}\oint_A x{dx\over y}.
$$
The period integral is evaluated in the Weierstrass standard form as 
$$
\gamma_0 =  -{1\over g_0}\oint_A (t+{u\over 3}){dt\over 
y}  = {1\over 3}({\pi\over g_0})^2E_2(\tau)- {u\over 3}.
$$
According to (\ref{5.8})$\sim$(\ref{5.10}) the initial condition (\ref{a'}) can be put in 
the form 
\begin{eqnarray}
a = -\Lambda_1\oint_A (x-u){dx\over y}    \label{inicon}
\end{eqnarray}
in which the integrand is  the Seiberg-Witten differential $dS$. 
 The free energies (\ref{4.11}) and (\ref{4.12}) are given with all $\Lambda_M$ vanishing 
except $\Lambda_1$, {\it i.e.}, 
\begin{eqnarray}
{\cal F} &=& {\cal F}|_{\tilde T_1 = T_1-\Lambda_1, \tilde T_{M\ge 2} = T_{M\ge 2}}, 
\label{F}
\end{eqnarray}
and 
\begin{eqnarray}
{\cal F}_0(a) &=& {\cal F}|_{{\rm all}\ T_M =0} \nonumber\\
              &=& {1\over 2}a^2 \tau(a,\vec 0) -  a\Lambda_1 \tau_{01}(\tau(a,\vec 0)) + 
{1\over 2} \Lambda_1^2\tau_{11}(\tau(a,\vec 0)). \label{F0}
\end{eqnarray}
Here note that $\tilde\tau_{MN}$ is reduced to $\tau_{MN}$ because $\Delta_M(x,\tau) =0$. 
To get  explicit forms of the free energies (\ref{F}) and (\ref{F0}) 
we need to calculate all of $\tau_{MN}$ as  in Appendix B.
 With the free energy (\ref{F0}) the effective Lagrangian of the $N=2$   Yang-Mills theory 
with $SU(2)$ is given by\cite{SW}  
\begin{eqnarray}
 W_{SW} = {1\over 4\pi}{\rm Im}[\int d^4\theta\ {\partial {\cal F}_0(A)\over \partial 
A}\bar A
 + \int d^2\theta\ {1\over 2}\tau(A,\vec 0) W_\alpha W^\alpha ],  \nonumber
\end{eqnarray}
in which  $A$ and $W_\alpha$ are  $ N=1\ U(1) $ chiral and vector superfields respectively.

\vspace{0.5cm}

\subsection{$N=1^*$ theory }

\noindent
 The $N=1^*$ theory with $U(N)$ is characterized by the effective 
superpotential\cite{DV2,DV3}
\begin{eqnarray}
W_{eff} = \int d^2\theta\ (Na_D - {\mathop{\tau}^\circ} a).   \label{eff}
\end{eqnarray}
Here $a$ and $a_D$ are given by
\begin{eqnarray}
a &=&  {\pi\over 72}( E_2(\tau)^2 - E_4(\tau) ), \label{a} \\
a_D &=&  \tau{\pi\over 72}( E_2(\tau)^2 - E_4(\tau) )
 + {i\over 12}E_2(\tau) = {\p {\cal F}_0 \over \p a},  \label{ad}
\end{eqnarray}
which were denoted by $\Pi_A (=2\pi i S) $ and $\Pi_B$ respectively in \cite{DV2,DV3}. It 
is important to 
remark that ${\p a_D \over \p a} = \tau$ is guaranteed by the formula
\begin{eqnarray}
{\partial \over \partial \tau} E_2(\tau) = {i\pi \over 6}( E_2(\tau)^2 - E_4(\tau)). 
\label{formula}
\end{eqnarray}
The bare coupling $\displaystyle{\mathop{\tau}^\circ} (\equiv \theta/2\pi+4\pi/g_{eff}^2) $ 
is related by extremizing the superpotential as\cite{DV2}
$$
\tau = {\displaystyle{\mathop{\tau}^\circ} + k \over N}, 
\quad\quad\quad k = 0,1,2,\cdots, N-1.
$$
We shall find an explicit form of the free energy  ${\cal F}_0$ in (\ref{ad}). 
To this end, we shall put the $N=1^*$ theory in the formalism developed in Sections 5 and 
6. Note the similarity between  the set of the equations (\ref{a})$\sim$(\ref{formula}) and 
that of  (\ref{4.5}),  (\ref{5.10'}) and 
\begin{eqnarray} 
{\p \over \p \tau}\sum_{M=1}^{M_0} \Lambda_M \tau_{0M} = \sum_{M=1}^{M_0} \Lambda_M 
s_M(\tau).  \label{kk}
\end{eqnarray}
The last equation is due to (\ref{3.2}) and(\ref{3.1''}). 
Therefore if $a$, given by (\ref{a}), is identified as 
\begin{eqnarray}
{\pi\over 72}(E_2(\tau)^2 - E_4(\tau)) = \sum_{M=1}^{M_0}\Lambda_M s_M(\tau),
 \label{aa}
\end{eqnarray}
with certain parameters $(\Lambda_1,\Lambda_2,\cdots,\Lambda_{M_{0}})$,  we have in 
(\ref{ad}) 
\begin{eqnarray}
 {i\over 12}E_2(\tau)  = -\sum_{M=1}^{M_0} \Lambda_M \tau_{0M}, \label{con}
\end{eqnarray}
and {\it vice versa}.  (\ref{a}) can be considered as the constraint which gives by 
inversion  the initial condition $\tau_0(a)$ for the hodograph solution. 
 Then the identification (\ref{aa}) or (\ref{con}) implies that 
the $N=1^*$ theory can be characterized by the differential $dS$, (\ref{5.10}) and 
correspondingly 
the free energy ${\cal F}_0$, (\ref{4.12}). In the present case they read respectively 
\begin{eqnarray}
dS = -\sum_{M=1}^{M_0} \Lambda_M d\tilde\Omega_M
 +{\pi\over 72}(E_2(\tau)^2 - E_4(\tau)  )d\omega_0,  \label{dS1'}
\end{eqnarray}
and 
\begin{eqnarray}
{\cal F}_0(a) = {1\over 2}a^2 \tau(a,\vec 0)
+{i\over 12}a E_2(\tau(a,\vec 0)) + {1\over 2}\sum_{M,N=1}^{M_0} \Lambda_M\Lambda_N 
\tilde\tau_{MN}(\tau(a,\vec 0)).  \label{FFF}
\end{eqnarray}
Calculating $Q_M(x)$ and $\tilde\tau_{MN}$ by using Appendix B  we find a concrete form of 
the free energy  ${\cal F}_0$ in (\ref{ad}). 
It is important remark that this free energy remains the same independently of which set to 
take for $(\Lambda_1,\Lambda_2,\cdots,\Lambda_{M_{0}})$. It is due to  (\ref{aa}) or 
equivalently (\ref{con}),  as shown by  the following calculation
\begin{eqnarray}
{\p \over \p \tau}\sum_{M,N=1}^{M_0} &\Lambda_M&\Lambda_N \tilde\tau_{MN}  \nonumber \\
 &=& \sum_{M,N=1}^{M_0}\Lambda_M\Lambda_N [{\p \over \p \tau}\tau_{MN} + 
   2\pi i {\rm res}_{\xi=0}[ \xi^{-2M}d\xi\Delta_N(x,\tau) ] ]      \nonumber 
\\
  &=& -2\pi i \sum_{M,N=1}^{M_0}\Lambda_M\Lambda_N{\rm 
res}_{\xi=0}[{\scriptstyle{\xi^{-2M+1}}\over \scriptstyle{2M-1}}s_N(\tau){\p \over \p \tau}
d\omega_0] 
    \label{cal} \\
   &=& [\sum_{M=1}^{M_0}\Lambda_M s_M(\tau)]^2 = [{\pi\over 72}(E_2(\tau)^2 - 
E_4(\tau))]^2,  \nonumber
\end{eqnarray}
with the help of (\ref{2.13}), (\ref{whit}), (\ref{tildetau}) and (\ref{kk}). In other 
words,  depending 
on the choice of $(\Lambda_1,\Lambda_2,\cdots,\Lambda_{M_{0}})$ we are here changing the 
characteristic speeds $s_M(\tau)$, {\it i.e.},  parameterization of the  curve, so as to 
satisfy (\ref{aa}) or equivalently (\ref{con}). On the  contrary, 
 in Section 5 we have done in the opposite way, {\it i.e.}, we  kept the characteristic 
speeds the same, but changed the initial condition $\tau_0(a)$ by choosing 
$(\Lambda_1,\Lambda_2,\cdots,\Lambda_{M_{0}})$ differently.

\vspace{0.5cm}

To see this, we closely look into  the constraint (\ref{con}). 
 By using the formulae (\ref{2point})  for $\tau_{0M}$ it reads
\begin{eqnarray}
\Lambda_1 = -{1\over 24}{g_0 \over \pi}E_2(\tau),   \quad\quad {\rm for}\ M_0 = 1, 
\label{con1} \\    
\Lambda_2 c + 6\Lambda_1 = -{1\over 4}{g_0\over \pi}E_2(\tau), \quad\quad {\rm for}\ M_0 = 
2,   \label{con2} \\
 \nonumber \\
\Lambda_3 c^2 + 4\Lambda_2 c + {3\over 5}\Lambda_3 g_2 + 24\Lambda_1 = 
-{g_0\over \pi}E_2(\tau), \quad\quad {\rm for}\ M_0 = 
3,   \label{con3}
\end{eqnarray}
and so on.

\vspace{0.5cm}

\noindent
i) \underline {  case with $M_0=1$}

  It suffices to consider the constraint (\ref{con1})  as the ``gauge" condition for $g_0$, 
discussed in Section 3.  The differential (\ref{dS1'}) takes the form
\begin{eqnarray}
dS &=& - \Lambda_1 [(x-{c\over 3}- g_1){dx\over y}-d(\int^\tau d\tau \Delta_1(x,\tau))]    
\nonumber \\
&\quad& \hspace{1cm}
 + {\pi\over 72}(E_2(\tau)^2 - E_4(\tau)  
)d\omega_0,\label{dSS}
\end{eqnarray}
by determining $Q_1(x)$ by the requirement (\ref{2.8}). 
The corresponding free energy (\ref{FFF}) becomes
\begin{eqnarray}
{\cal F}_0(a) = {1\over 2}a^2 \tau(a,\vec 0)
+{i\over 12}a E_2(\tau(a,\vec 0)) + {1\over 2}\Lambda_1^2 \tilde\tau_{11}
(\tau(a,\vec 0)),  \label{x1}
\end{eqnarray}
with $\tilde\tau_{11}$ defined by (\ref{tildetau}), {\it i.e.},
\begin{eqnarray}
\tilde\tau_{11} = 2\pi i[{c\over 6} - g_1 +
   \int^\tau d\tau\  {\rm res}_{\xi=0}[\xi^{-2}d\xi \Delta_1(x,\tau)]].  \label{11}
\end{eqnarray}
Here $\tau_{11}$ was calculated by (\ref{2point}).

\vspace{0.5cm}

\noindent
ii) \underline{case with $M_0=2$}

For simplicity we set  $\Lambda_1 =0$. Then (\ref{con2}) requires  $c$ to be 
\begin{eqnarray}
c= -{1\over 4\Lambda_2}{g_0\over \pi}E_2(\tau).  \label{M2}
\end{eqnarray}
Here $g_0$ is still to be fixed as the ``gauge" condition.  With $c$ constrained as such, 
the differential (\ref{dS1'}) and the free energy (\ref{FFF}) are 
 respectively given by 
\begin{eqnarray}
dS &=& - \Lambda_2 [(x^2 -{c\over 2}x +{c^2\over 18} -{c\over 6}g_1-{g_2\over 12}){dx\over 
y} -d(\int^\tau d\tau \Delta_2(x,\tau))] \nonumber \\
&\quad& \hspace{2cm} + {\pi\over 72}(E_2(\tau)^2 - E_4(\tau)  )d\omega_0, \label{dS2}
\end{eqnarray}
and 
\begin{eqnarray}
{\cal F}_0(a) = {1\over 2}a^2 \tau(a,\vec 0) 
+{i\over 12}a E_2(\tau(a,\vec 0) ) + {1\over 2}\Lambda_2^2 \tilde\tau_{22}
(\tau(a,\vec 0)),  \label{x2}
\end{eqnarray}
with
\begin{eqnarray}
\tilde\tau_{22} =  \pi i[{ c^3\over 324}  &-& {c^2 \over 18}g_1 + {c\over 36} g_2 + {1\over 
12}g_3   \nonumber \\
 &+& 2\int^\tau d\tau\  {\rm res}_{\xi=0}[\xi^{-4}d\xi \Delta_2(x,\tau)]].    \label{22}
\end{eqnarray}
Here $Q_2(x)$ in $dS$ was calculated by means of (\ref{A.4}) and (\ref{gamma00}), while  
$\tilde\tau_{22}$ in (\ref{2point}).

\vspace{0.5cm}

\noindent
iii) \underline{case with $M_0=3$}:

 For simplicity set  $\Lambda_2 =\Lambda_1 = 0$.  Then (\ref{con3}) constrains $c$ such 
that
\begin{eqnarray}
c^2 = -{1\over \Lambda_3}[{g_0\over \pi}E_2(\tau)+{3\over 5}g_2]. \label{M3}
\end{eqnarray}
Here also  the ``gauge" freedom $g_0$ is still to be fixed. 
 Similarly to the previous case we calculate
the differential (\ref{dS1'}) and the free energy  (\ref{FFF}) by the formulae in Appendix 
B. Then with $c$ given by (\ref{M3}) they are found 
to take the respective forms  
\begin{eqnarray}
dS &=& -\Lambda_3 [\{x^3 -{c\over 2}x^2 +({c^2\over 24} -{g_2\over 8})x + {c^3\over 216} 
-{c^2\over 24}g_1 - {g_3\over 10} -{g_1 g_2\over 40}\}{dx\over y} \nonumber \\
&\quad& \hspace{1cm} - d(\int^\tau d\tau \Delta_3(x,\tau))]  
+ {\pi\over 72}(E_2(\tau)^2 - E_4(\tau)  )d\omega_0,   \label{dS3}
\end{eqnarray}
and 
\begin{eqnarray}
{\cal F}_0(a)
    = {1\over 2}a^2 \tau(a,\vec 0) + {i\over 12} a E_2(\tau(a,\vec 0)) + {1\over 
2}\Lambda_3^2 \tilde \tau_{33}(\tau(a,\vec 0)), \label{F2}
\end{eqnarray}
with
\begin{eqnarray}
\tilde\tau_{33} &=& \pi i[ {c^5\over 8640} - {c^4\over 288}g_1 + {c^3\over 288}g_2 + 
{c^2\over 240}(-g_2 g_1 + 6g_3)  \nonumber \\
&\quad& 
\hspace{1cm} + {c\over 192}(g_2)^2 +
{3\over 400}g_3 g_2  - {1\over 800}(g_2)^2 g_1  \nonumber \\
  &\quad& \hspace{2cm} +  2\int^\tau d\tau\  {\rm res}_{\xi=0}[\xi^{-6}d\xi 
\Delta_3(x,\tau)]  
].  \label{33}
\end{eqnarray}

\vspace{0.5cm}

 So far there remains the ``gauge" freedom in both cases ii) and iii). For the case ii) let 
us take the ``gauge"  $g_0 = 2\pi$ and  set $\Lambda_2 =1$.  Then  the constraint 
(\ref{M2}) becomes  $c = -{1\over 2}E_2(\tau)$. The Dijkgraaf-Vafa differential (\ref{dS2}) 
is simplified as
\begin{eqnarray} 
dS = -[t^2 - {1\over 12} E_2(\tau)t - {1\over 72}E_2(\tau)^2]{dt\over y}
 +d(\int^\tau d\tau \Delta_2(x,\tau)). \label{DV}
\end{eqnarray}
with $t= x +{1\over 6}E_2(\tau)$. 
This is the Dijkgraaf-Vafa differential given in  \cite{DV2}. 
The free energy is reduced to 
\begin{eqnarray}
{\cal F}_0(a) &=& {1\over 2}a^2 \tau(a,\vec 0) + 
{i\over 12}a E_2(\tau(a,\vec 0))   \nonumber \\
 &\quad& + {\pi i\over 2}[{1 \over 1296}E_2(\tau(a,\vec 0))^3 - {1\over 864}E_2(\tau(a,\vec 
0))E_4(\tau(a,\vec 0)) + {1\over 2592}E_6(\tau(a,\vec 0))]. \nonumber \\
  & &  \nonumber \\
 &\quad&  + \pi i\int^\tau d\tau\  {\rm res}_{\xi=0}[\xi^{-4}d\xi \Delta_2(x,\tau)].  
\nonumber
\end{eqnarray}
By plugging $a$ given by (\ref{a}) and evaluating the last term  as (\ref{res}) in Appendix 
C it takes the form
\begin{eqnarray}
 {\cal F}_0(a) &=& {1\over 2}({\pi \over 72})^2(E_2(\tau)^2 - E_4(\tau))^2 \tau 
+ {\pi i\over 72^2}(8 E_2(\tau)^3 - 9E_2(\tau)E_4(\tau) + E_6(\tau)]. 
\nonumber \\
  & &              \nonumber\\
 &\quad&  + {1\over 2}({\pi\over 72})^2 \int^\tau d\tau[ 6 E_2(\tau)E_6(\tau) 
 -9 E_2^2(\tau)E_4(\tau) + 3E_2(\tau)^4],   \nonumber
\end{eqnarray}
which may be now considered as  a function of $\tau$.

\vspace{0.5cm}

Thus we have shown that the $N=1^*$ theory can be characterized by the different  DV 
differentials (\ref{dSS}), (\ref{dS2}) and (\ref{dS3}). But the corresponding free energies 
(\ref{x1}), (\ref{x2}) and (\ref{F2}) are all kept the same, 
 as has been  proved by the calculation (\ref{cal}). Nonetheless they  look quite 
different. As a consistency check of the whole formalism, it is worth showing  directly 
that 
\begin{eqnarray}
\Lambda_1^2{\p \over \p\tau }\tilde\tau_{11} = \Lambda_2^2{\p \over \p\tau }\tilde\tau_{22} 
= \Lambda_3^2{\p \over \p\tau }\tilde\tau_{33}  =  [{\pi\over 72}(E_2(\tau)^2 - E_4(\tau))]
^2  \label{diftau}
\end{eqnarray}
for  $\tilde\tau_{11},  \tilde\tau_{22}$ and $\tilde\tau_{33}$ given by (\ref{11}), 
(\ref{22}) and (\ref{33}). It will be done in Appendix C. 

\vspace{0.5cm}

But difference due to the respective characterization of the $N=1^*$ theory by
 (\ref{dSS}), (\ref{dS2}) and (\ref{dS3}) appears when we discuss the Whitham deformation 
of the theory. The Whitham deformation is governed by the hodograph solution (\ref{4.6}). 
 The curve is parameterized in such a way that for each case of i), ii) and iii)  one of 
the 
characteristic speeds $s_M(\tau), M=1,2,\cdots$, takes the fixed functional form, {\it 
i.e.},  
\begin{eqnarray}
\left. \begin{array}{rl}
   \Lambda_1 s_1(\tau) \ ( {\rm \ of\ case \  i)})  \\
  \Lambda_2 s_2(\tau) \ ( {\rm \ of\ case \ ii)})  \\
 \Lambda_3 s_3(\tau) \ ( {\rm \ of\ case \ iii)})
\end{array}\right\} 
= {\pi \over 72}(E_2(\tau)^2 - E_4(\tau)).   \label{initial}
\end{eqnarray}
Then all other $s_M(\tau)$ differ depending on the case to be considered. Hence  the 
hodograph solution (\ref{4.6}) flows  differently from the same initial condition 
$\tau_0(a)$. So does the free energy ${\cal F}_0(a)$, as   given by (\ref{4.11}).  ( See  
Fig. 1.) 
In other words,   the curve (\ref{2.1}) is parameterized  differently for each case  of i), 
ii) and iii). The branch points  are given  by (\ref{C.2})  as 
\begin{eqnarray}
 u(\tau) &=& {c\over 3} + {1\over 3}({\pi\over g_0})^2[\theta_3(\tau)^4 + \theta_0(\tau)^4 
], \nonumber \\
v(\tau) &=& {c\over 3} + {1\over 3}({\pi\over g_0})^2[\theta_2(\tau)^4 - \theta_0(\tau)^4 
],  \\
w(\tau) &=&  {c\over 3} + {1\over 3}({\pi\over g_0})^2[\theta_2(\tau)^4 + \theta_3(\tau)^4 
], \nonumber 
\end{eqnarray} \FIGURE{\epsfig{file=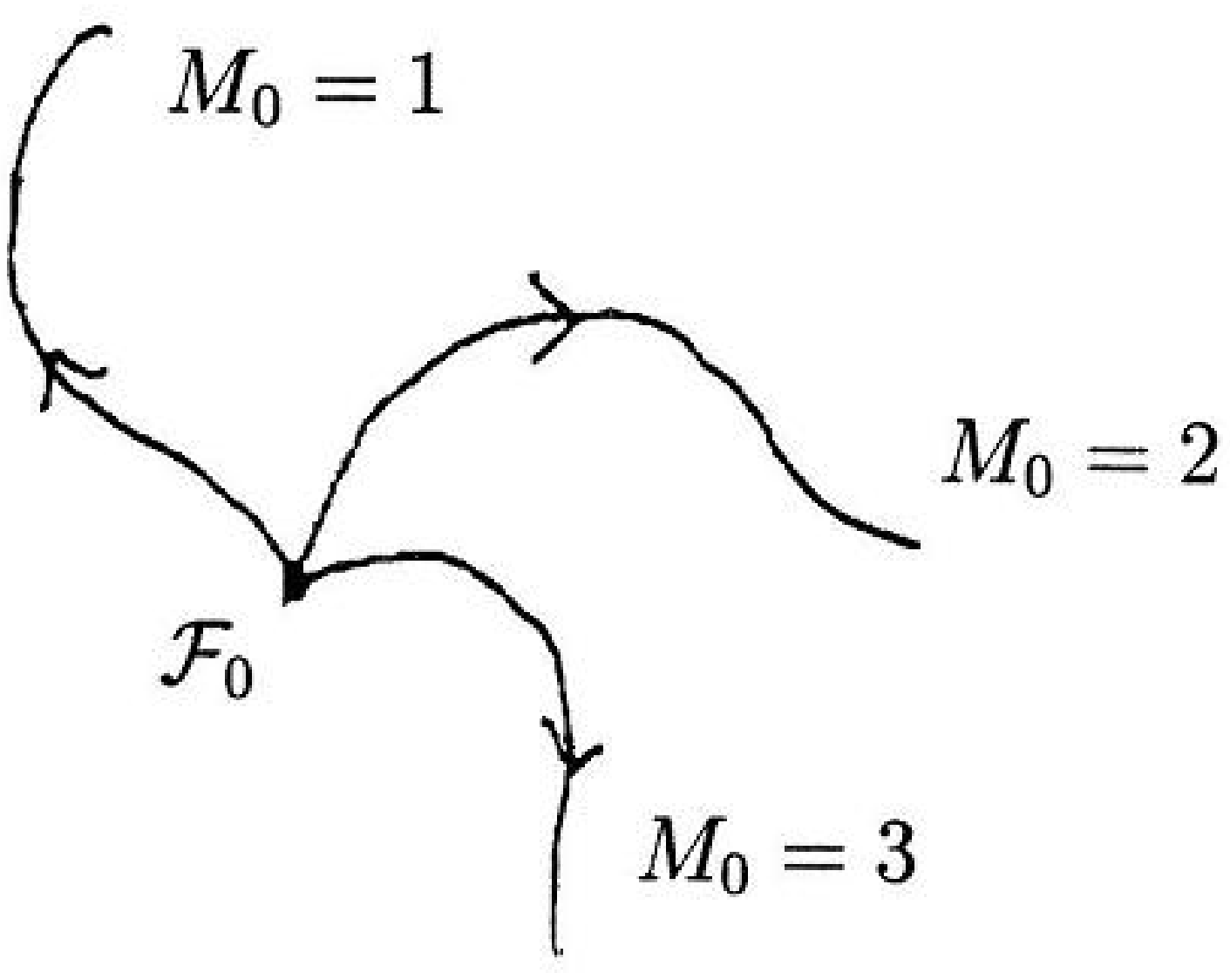, width=4cm}
\caption{Flows of the free energy ${\cal F}_0$}}  

\noindent
in which $c$ as well as $g_0$  are different  depending on the case. ( $g_0$ is  the 
``gauge" 
freedom to be fixed arbitrarily, except for case i)). 
 Then these branch points  move through deformation of $\tau$  as given by the hodograpgh 
solution (\ref{4.6}). The curve  might happen to degenerate at some flow points $\vec T 
\ne 0$. We may expect a large variety of degeneracy of the $N=1^*$ theory,  depending  on 
the choice of $c$, though the free energy is initially the same. 

\vspace{1cm}

\section{Interpretation of the flow parameters $\Lambda_M$}
\setcounter{equation}{0}

We shall give a physical interpretation of the flow parameters $\Lambda_M$ which are so far 
mathematical. Namely we will show 
 that the free energy (\ref{4.12}) in the {\it small phase space} is equivalent to the one 
of the matrix model given by 
\begin{eqnarray}
Z = \int d\Phi {e^{-Ntr V(\Phi)} \over det([\Phi,-] + i)},  \label{N=1}
\end{eqnarray}
with 
$$
V(\Phi) = \sum_{M=0}^{M_0}\bar \Lambda_M \Phi^{2M}.
$$
Here $\Phi$ is an $N\times N$ hermite matrix and $\bar \Lambda_M $ is a linear combination 
of  $\Lambda_M $ which will be given later. It can be done by making use of  the arguments 
in \cite{Mir}. Namely they discussed the relation between the Toda hierarchy and a matrix 
model. The matrix model (\ref{N=1}) is its variant and was studied in \cite{DV2,DV3, 
Dorey}. There underlies the KdV hierarchy.
 So we shall adapt the arguments in \cite{Mir} to this case.  
By diagonalizing  $\Phi$ the integral (\ref{N=1}) is reduced to an integral over  the 
eigenvalues  $\lambda_1,\lambda_2,\cdots,\lambda_N$.  Then we obtain 
the saddle point equation 
\begin{eqnarray}
-NV'(\lambda_I) + \sum_{J\ne I}[ {2\over \lambda_I -\lambda_J} -
{1 \over \lambda_I -\lambda_J +i}-{1 \over \lambda_I -\lambda_J -i}]
=0.  \label{saddle}
\end{eqnarray}
Let us take the large $N$ limit and assume that the eigenvalues spread out along the real 
line $[-\alpha,\alpha]$ with the density $\rho(\lambda)$. 
The integral (\ref{N=1})  is written in the form
\begin{eqnarray}
\log Z &=& -N^2 \{ \int_{-\alpha}^\alpha d\lambda \rho(\lambda)V(\lambda) 
- {1\over 2} \int_{-\alpha}^\alpha d\lambda \rho(\lambda) \int_{-\alpha}^\alpha 
d\lambda'\rho(\lambda') \nonumber \\
& \quad &\cdot[ 2\log (\lambda -\lambda') -  \log (\lambda -\lambda' +i)- \log (\lambda 
-\lambda' -i) ]\}. \label{logZ}
\end{eqnarray} 
and (\ref{saddle}) becomes 
\begin{eqnarray}
V'(\lambda) - \int_{-\alpha}^\alpha d\lambda' \rho(\lambda')
 [ {2\over \lambda -\lambda'} - {1\over \lambda -\lambda' +i}- {1\over \lambda -\lambda' 
-i} ] =0, 
\label{saddle2}
\end{eqnarray}
for $ \lambda \in [-\alpha,\alpha] $. The {\it l.h.s.} of this equation is a force acting 
on a test eigenvalue $\lambda$. Introducing the resolvent
\begin{eqnarray}
\omega(\lambda) = \int_{-\alpha}^\alpha d\lambda' { \rho(\lambda')\over \lambda -\lambda'}, 
\nonumber
\end{eqnarray}
we rewrite this force as 
\begin{eqnarray}
f(\lambda)\equiv V'(\lambda)\ -\ [2\omega(\lambda) \ -\ \omega(\lambda +i)\ -\ 
\omega(\lambda -i)].   \label{force}
\end{eqnarray}
Then the saddle point equation (\ref{saddle2}) reads 
\begin{eqnarray}
f(\lambda) = 0, \quad\quad\quad {\rm for} \quad \lambda \in [-\alpha,\alpha]
. \label{saddle3} 
\end{eqnarray}
We further introduce the function 
\begin{eqnarray}
G(\lambda) = U(\lambda) + i[\omega(\lambda +{i\over 2})-\omega(\lambda -{i\over 2})], 
\label{G}
\end{eqnarray}
with a polynomial $U(\lambda)$ such that 
\begin{eqnarray}
V'(\lambda) = -i[ U(\lambda+{i\over 2}) -U(\lambda-{i\over 2})]. \label{defU}
\end{eqnarray}
To be concrete, $U(\lambda)$ is found in the form 
\begin{eqnarray}
{\partial U(\lambda)  \over \partial \bar\Lambda_M }
 = \sum_{k=0}^M c_k^{(M)} \lambda^{2k}, \quad\quad\quad c_M^{(M)} =1,
\label{singular}
\end{eqnarray}
in which   $c_k^{(M)}$ are numerical values recursively determined  with the requirement 
(\ref{defU}). In terms of the function $G(\lambda)$ 
the saddle point equation (\ref{saddle3}) becomes
\begin{eqnarray}
G(\lambda+{i\over 2}) = G(\lambda-{i\over 2}), \quad\quad\quad {\rm for} \quad \lambda \in 
[-\alpha,\alpha]
. \label{saddle4}
\end{eqnarray}
This function has two branch cuts along 
$$
{\cal C}_+ = [ -\alpha +{i\over 2} , \alpha +{i\over 2}], \quad\quad\quad\quad
{\cal C}_- = [ -\alpha -{i\over 2} , \alpha -{i\over 2}].
$$
The saddle point equation (\ref{saddle4}) implies that $G(\lambda)d\lambda$ is a 
meromorphic differential on an elliptic curve with the pole structure read in (\ref{G}). We 
take $A$-cycle and the dual $B$-cycle  in the same way as in \cite{DV2}. 
We also define  the function $\xi(\lambda)$  by integrating the force (\ref{force}):
\begin{eqnarray}
\xi(\lambda) &\equiv& \int_{\infty}^\lambda d\lambda f(\lambda) 
\label{defxi} \\
 &=& V(\lambda) - 
\int_{-\alpha}^\alpha d\lambda'\rho(\lambda') 
[ 2\log (\lambda -\lambda') -  \log (\lambda -\lambda' +i)- \log (\lambda -\lambda' -i) ].   
\nonumber
\end{eqnarray}
 It is known \cite{DV2} that for $ \lambda \in [-\alpha,\alpha] $ this function  is  
$\lambda$-independent as 
\begin{eqnarray}
\xi(\lambda) = \int_{\infty}^\alpha d\lambda f(\lambda) 
=i\oint_B d\lambda  G(\lambda). \label{xi1}
\end{eqnarray}

\vspace{0.5cm}

Noting that the jump in $G(\lambda)$ along  the branch cut ${\cal C}_+$ (going upwards ) is 
$2\pi \rho(\lambda)$, we rewrite the free energy (\ref{logZ}) as
\begin{eqnarray}
\log Z &=& -N^2\{ {1\over 2} \oint_A {d\lambda \over 2\pi }
G(\lambda)V(\lambda) \ - \ 
{1\over 8}\oint_A {d\lambda \over 2\pi }\oint_A {d\lambda' \over 2\pi }
G(\lambda)G(\lambda')    \nonumber \\ 
&\quad&  
\cdot[ 2\log (\lambda -\lambda') -  \log (\lambda -\lambda' +i)- \log (\lambda -\lambda' 
-i) ]\}.   \label{logZ'}
\end{eqnarray}
Let us define the variable $\bar a$  by
\begin{eqnarray}
\bar a = \oint_A G(\lambda) d\lambda.  \label{defa}
\end{eqnarray}
We differentiate the free energy (\ref{logZ'}) by this $\bar a$:
\begin{eqnarray}
{\partial \log Z \over \partial \bar a} = -{N^2\over 2}  \oint_A {d\lambda \over 2\pi 
}{\partial G(\lambda) \over \partial \bar a }\xi(\lambda).    \nonumber 
\end{eqnarray}
Remember the fact that $\xi(\lambda)$ is 
$\lambda$-independent on the $A$-cycle and written as (\ref{xi1}). Then this 
 becomes 
\begin{eqnarray}
 {\partial \log Z \over \partial \bar a} &=& -i{N^2\over 2}  \oint_A {d\lambda \over 2\pi }
{\partial G(\lambda) \over \partial \bar a }\oint_B d\lambda G(\lambda)
  \nonumber \\
&=& -i{N^2\over 4\pi} \oint_B d\lambda G(\lambda),   \label{difflogZ1}
\end{eqnarray}
due to  (\ref{defa}). By a similar calculation we have 
\begin{eqnarray}
{\partial \log Z \over \partial \bar\Lambda_M } = 
 -N^2[ {1\over 2} \oint_A {d\lambda \over 2\pi }{\partial G(\lambda) \over \partial 
\bar\Lambda_M }\xi(\lambda) + {1\over 2} \oint_A {d\lambda \over 2\pi } G(\lambda){\partial 
V(\lambda) \over  \partial \bar\Lambda_M } ]. 
\end{eqnarray}
Use  ${\partial \bar a \over \partial \bar \Lambda_M}=0$ and the properties  of 
$\xi(\lambda)$ 
again. It may be written as 
\begin{eqnarray}
{\partial \log Z \over \partial \bar\Lambda_M } =  
-i{N^2\over 2}  \oint_A {d\lambda \over 2\pi } G(\lambda) 
\oint_B d\lambda {\partial G(\lambda)\over \partial \bar\Lambda_M}.
\end{eqnarray}
Finally we have recourse  to the Riemann bilinear formula to proceed the calculation. 
Knowing the singular behavior of  ${\partial G(\lambda)\over \partial \bar\Lambda_M} $ from 
(\ref{singular}) 
we obtain 
\begin{eqnarray}
{\partial \log Z \over \partial \bar\Lambda_M } = 
i {N^2\over 4\pi } \sum_{k=0}^M  
 2\pi i\ c_k^{(M)}  {\rm res}_{\lambda =\infty} 
[{\scriptstyle{\lambda^{2k+1}}\over {\scriptstyle{2k+1}}} G(\lambda)d\lambda ].  
\label{difflogZ2}
\end{eqnarray}

\vspace{0.5cm}

We think of the mapping from the $\lambda$-plane to the $z$-plane 
parameterizing  the torus with the period ${\omega_2 \over \omega_1}$. 
In \cite{Dorey} it was given by 
\begin{eqnarray}
\lambda(z) = -{\omega_1\over \pi}[ \zeta(z) - {\zeta(\omega_1)\over \omega_1} z ] = 
-{\omega_1\over \pi } [{1\over z} + O(z)].  \label{zeta}
\end{eqnarray}
with  $\zeta(z)$ the Weierstrass $\zeta$-function. 
By using this coordinate $z$  (\ref{difflogZ2}) becomes
\begin{eqnarray}
{\partial \log Z \over \partial \bar\Lambda_M } = 
-i {N^2\over 4\pi } \sum_{k=0}^M  
 2\pi i\ c_k^{(M)}({\omega_1\over \pi })^{2k+1}  {\rm res}_{z =0} 
[{\scriptstyle{z^{-2k-1}}\over {\scriptstyle{2k+1}}} G(\lambda)d\lambda ].  
\label{difflogZ3}
\end{eqnarray}
If we can identify  the meromorphic differential $G(\lambda)d\lambda$ with $dS$ and $\log  
Z$ with $-i{\cal F}_0$,  
then 
(\ref{difflogZ1}) and (\ref{difflogZ2}) become the equations (\ref{formula1}) and 
(\ref{formula2}) by  appropriate linear combination of $\bar\Lambda_M$. Thus we can 
interpret the flow parameters $\Lambda_M$ of the Whitham deformation as the coupling 
constants $\bar\Lambda_M$ of the  matrix model (\ref{N=1}). So we are left with the task to 
show $G(\lambda)d\lambda$ to take the same form as (\ref{5.10}). 
To this end we write the elliptic curve (\ref{2.1}) in the form 
\begin{eqnarray}
y^2 = 4t^3 -g_2\, t -g_3,
\end{eqnarray}
and note that it is mapped to the $z$-plane through the Weierstrass ${\cal P}$-function 
\begin{eqnarray}
t = -\zeta'(z) = {\cal P}(z), \quad \quad\quad y= {\cal P}'(z). \label{P}
\end{eqnarray}
Following \cite{Dorey} we shall  determine 
the elliptic function $G(\lambda)$. From (\ref{G}), ({\ref{singular}) and (\ref{zeta}) we 
know the pole singularity 
\begin{eqnarray}
[ G(\lambda(z)) ]_- = 
  \sum_{M=0}^{M_0}\bar\Lambda_M\sum_{k=0}^{M}c_k^{(M)} ({\omega_1\over \pi})^{2k}
 \sum_{i=0}^{k} d_i^{(k)} z^{-2i},  
\end{eqnarray}
in which $[\cdots ]_-$ indicates the part of non-positive powers in $z$ and 
$d_i^{(k)}$ are calculable constants with $d_{k}^{(k)} = 1$. 
By using the expansion 
\begin{eqnarray}
{\cal P}(z) = {1\over z^2} + {g_2\over 20}z^2 + {g_3\over 28}z^4 + \cdots ,
\end{eqnarray}
it may be expressed by a polynomial of ${\cal P}(z)$ 
That is, 
\begin{eqnarray}
[ G(\lambda(z)) ]_- &=& 
\sum_{M=0}^{M_0}\bar\Lambda_M\sum_{k=0}^{M}c_k^{(M)} ({\omega_1\over \pi})^{2k}
 [{\cal P}(z)^{k} + d_{k-1}^{(k)}{\cal P}(z)^{k-1} + \cdots ]_-  \nonumber \\
 &\equiv & [\sum_{M=0}^{M_0}\bar\Lambda_M\sum_{k=0}^{M}c_k^{(M)} ({\omega_1\over \pi})^{2k}
\bar P_{k}({\cal P}(z)) ]_-,
\end{eqnarray}
with a polynomial $\bar P_{k}(\cdot)$ of degree $k$. Since an elliptic function is 
determined uniquely by the pole singularity, so that 
\begin{eqnarray}
G(\lambda(z))  =
  \sum_{M=0}^{M_0}\bar\Lambda_M\sum_{k=0}^{M}c_k^{(M)} ({\omega_1\over \pi})^{2k}\bar 
P_{k}({\cal P}(z)).   \label{Gfinal}
\end{eqnarray}
By means of (\ref{zeta}), (\ref{P}) and (\ref{Gfinal})  the meromorphic differential 
$G(\lambda)d\lambda$ can be written as
\begin{eqnarray}
G(\lambda)d\lambda = \sum_{M=0}^{M_0}\bar\Lambda_M\sum_{k=0}^{M}c_k^{(M)} ({\omega_1\over 
\pi})^{2k+1}[t + {\zeta(\omega_1)\over \omega_1} ]\bar P_{k}
(t){dt\over y}.
\end{eqnarray}
By a linear transformation from $\bar\Lambda_M$ to $\Lambda_M$ this takes the form (\ref{5.10}), except for the boundary term in $d\tilde \Omega_M$ which vanishes at $t=\infty$. But, 
when we interpreted  the saddle point equation (\ref{saddle4}),  we could allow $G(\lambda)
d\lambda$ to have a boundary term like $d(\cdots )$ which is not meromorphic. As long as it 
vanishes  at $\lambda=\infty$, the argument thereafter goes through without any 
modification. Appropriately fixing this freedom would yield the boundary term in $dS$. Or 
note that this boundary term depends on the gauge fixing discussed in Section 3. Then we 
can simply say that $\log Z$ of the matrix model corresponds to the free energy ${\cal 
F}_0$ of the Whitham hierarchy calculated  with the  particular gauge in which the boundary 
term disappears, for instance, $v= const$ and $w=const$.

\vspace{0.5cm}

Thus we were able to identify the free energy (\ref{4.12}) in the small phase space with 
the one of the matrix model (\ref{N=1}). 
The free energy (\ref{4.11}) in the {\it large phase space} may be identified as the one 
obtained by peturbing  the  model  (\ref{N=1}):
\begin{eqnarray}
Z = \int d\Phi\ {e^{-N tr[ \sum_{M=0}^{M_0} \bar\Lambda_M \Phi^{2M} -
 \sum_{M=0}^{\infty} \bar T_M   \Phi^{2M}]} \over det(\ [\Phi,-] \ \ + \ \ i\ )\ }.   
\label{pmodel}
\end{eqnarray}
In section 5 we have shown how to obtain the free energy (\ref{4.11}) by perturbing  the 
one (\ref{4.12}) in the {\it small phase space}.  We can obtain  the free energy of the  
model (\ref{pmodel}) by perturbing the model (\ref{N=1}) exactly  in the same  sense. 
 Thus the flow parameters $T_M, M=1,2,\cdots$ of the free energy (\ref{4.11}) in the {\it 
large phase space} can be interpreted as the coupling constants $\bar T_M$ of the 
perturbative interaction of the matrix model (\ref{pmodel}). 

\vspace{0.5cm}

The different characterizations of the free energy by (\ref{dSS}), (\ref{dS2}) and 
(\ref{dS3}) in the previous section corresponds to the matrix model (\ref{N=1}) with $M_0 = 
0,1,2$  after an appropriate setting of $\bar \Lambda_M$. The three  flows illustrated in 
Fig. 1 are interpreted as different perturbations of these matrix models by (\ref{pmodel}). 
Then 
(\ref{diftau}) or more concretely (\ref{initial}) is interpreted as
 a condition which sets 
 the free energy $\log Z$ with $M_0 = 0,1,2$ to be equal at $\bar T_M=0$. It can be done 
 by choosing the period $\tau = {\omega_2\over \omega_1}$ of the torus appropriately.

\vspace{0.5cm}

We have discussed only the $two$-cut solution of the matrix model (\ref{N=1}). We may 
consider  multi-cut solutions on a Riemann surface with higher genus. It would be 
interesting to extend the whole arguments in this section to such general cases.

\vspace{1cm}

\section{Conclusions}
\setcounter{equation}{0}

In this paper we have discussed the Whitham deformation of the free energy ${\cal F}_0$ 
which appears in the DV and SW theories. It amounts to discussing deformation of a relevant 
(hyper)elliptic curve with flow parameters. We were mainly concerned about the elliptic 
case. Generalization to the hyperelliptic case would be straightforward albeit with some 
technical complications. 
We then derived the Whitham equation for the period $\tau$. Its hodograph solution 
represents the Whitham deformation of the free energy ${\cal F}_0$ in the {\it large phase 
space} of flow parameters. To find a hodograph solution we have to impose a constraint  
which determines an initial functional form $\tau_0(a)$ or $a(\tau_0) (\equiv 
\tau_0^{-1}(a))$. It amounts to determining the free energy ${\cal F}_0$ in the  {\it small 
phase space}.  The main message of this paper is that  $\tau_0(a)$ 
and ${\cal F}_0$, given at one point in the {\it small phase space}, get deformed along 
different flows in the {\it large phase space}, when the elliptic curve is parameterized 
differently at that point.
 As an 
application of this argument we took the effective superpotential of the $N=1^*$ theory 
(\ref{eff}). We have shown that the same superpotential  can be indeed characterized  by 
the DV differentials (\ref{dSS}), (\ref{dS2}) and (\ref{dS3})
 on  different curves. 
  For each chosen DV differential the superpotential undergoes different Whitham 
deformations. We have also given an interpretation of these  Whitham deformations in terms 
of the matrix model. 

\vspace{0.5cm}

The free energy ${\cal F}_0$ of the $N=1^*$ theory took rather complicated 
forms for each case of i), ii) and iii). We emphasize that the term $\int ^\tau d\tau(...)$ 
coming from the boundary term in (\ref{whit}) is essential to have  the property $\tau = 
{\p^2 {\cal F}_0 \over \p a^2}$. If two of the branch points of the curve are fixed to be 
constant, for instance, $v=1$ and $w=-1$ as in the $N=2$ effective  Yang-Mills theory in 
Subsection 7.1, there is no contribution from the boundary. In such cases  the free energy 
${\cal F}_0$ is rather 
 simply calculated. But in general the boundary term was necessary for the compatibility of 
the Whitham deformation (\ref{3.1}) and (\ref{3.1'}).

\vspace{0.5cm}

The arguments in this paper can be applied for the SW theory in which the constraint  
(\ref{inicon}) is inverted to give  the initial condition
\begin{eqnarray}
\tau_0(a) = {g_{D0} \over g_0}\sim const. +   {i\over \pi } \log a^2, \quad\quad\quad a\sim 
\infty,   \label{swcon}
\end{eqnarray}
by using (\ref{C.4'}) and (\ref{C.4''}). 
This condition is essentially related to the effective coupling $g_{eff}$ of QCD. We might 
parameterize  the elliptic curve in the general form (\ref{gcurve}) and realize the initial 
constraint (\ref{swcon}) by the generalized differential (\ref{5.10}) with $M_0 \ge 2$. 
Then $a$ is expressed by the branch pints $u,v$ and $w$. They move obeying the Whitham 
equation. We might think of the Whitham equation as a master equation for studying  
degeneration of the curve, that is, critical phenomena of QCD.

\vspace{2cm}
\noindent

\acknowledgments
%{\Large\bf Acknowledgments}

\noindent
One of the authors (S.A) would like to thank Y. Kodama for useful discussions. 
His work was supported in part  by the Grant-in-Aid for Scientific Research No.
13135212.

\vspace{3cm}

\appendix

\section{The Weierstrass standard form and the period integrals}
%\apsect{A}

\setcounter{equation}{0}

We write an elliptic curve in the Weierstrass standard form
\begin{eqnarray}
y^2 = 4(t-e_1)(t-e_2)(t-e_3)   \label{C.1}
\end{eqnarray}
with 
\begin{eqnarray}
e_1 + e_2 + e_3 = 0.      \nonumber
\end{eqnarray}
Then the Jacobi $\theta$-functions are given in terms of the positions of the branch points 
$e_1,e_2$ and $e_3$:
\begin{eqnarray}
\theta_2(\tau) &=& \sum_{n=-\infty}^{\infty}q^{{1\over 2}(n-{1\over 2})^2}
  = i\left({2\omega_1\over \pi}\right)^{1\over 2}(e_2 - e_3)^{1\over 4},   \nonumber  \\
\theta_3(\tau) &=& \sum_{n=-\infty}^{\infty}q^{{1\over 2}n^2}
  = \left({2\omega_1\over \pi}\right)^{1\over 2}(e_1 - e_3)^{1\over 4},   \label{C.2}  \\
\theta_4(\tau) &=& \sum_{n=-\infty}^{\infty}(-1)^nq^{{1\over 2}n^2}
  = \left({2\omega_1\over \pi}\right)^{1\over 2}(e_1 - e_2)^{1\over 4},  \nonumber  
\end{eqnarray}
with 
\begin{eqnarray}
q &=& e^{2\pi i \tau},    \nonumber  \\
\omega_1 &=& \int_{-\infty}^{e_1} {dt \over \sqrt {4(t-e_1)(t-e_2)(t-e_3)}}.  \nonumber
\end{eqnarray}
(\ref{C.1}) may be also written in the form
\begin{eqnarray}
y^2 = 4t^3 -g_2t -g_3.
\end{eqnarray}
It is known that the coefficients are given by the Eisenstein series
\begin{eqnarray}
g_2 &=& 2(e_1^2 + e_2^2 + e_3^2 ) \nonumber \\
&=& {4\over 3}({\pi\over g_0})^4[1 + 240\sum_{n=1}^{\infty}{n^3q^n\over 1-q^n}] \equiv 
{4\over 3}({\pi\over g_0})^4 E_4(\tau) , \label{g2} \\
g_3 &=& 4e_1 e_2 e_3 \nonumber \\
 &=& {8\over 27}({\pi\over g_0})^6[1 - 504\sum_{n=1}^{\infty}{n^5q^n\over 1-q^n}] \equiv 
{8\over 27}({\pi\over g_0})^6 E_6(\tau).
\end{eqnarray}
with $g_0 = \oint_A {dt\over y} = -2\omega_1$. By using them we have  the following 
integration formulae
\begin{eqnarray}
{1\over g_0}\oint_A t{dt\over y} &=& g_1, \quad\quad\quad\quad\quad\quad
{1\over g_0}\oint_B t{dt\over y} = -\tau g_1 +{2\pi i\over g_0^2},
 \label{A1}  \\
{1\over g_0}\oint_A t^2{dt\over y} &=& {g_2\over 12}, 
\quad\quad\quad\quad\quad\quad\quad\quad
{1\over g_0}\oint_B t^2{dt\over y} = \tau{g_2\over 12}, \label{A2} \\
{1\over g_0}\oint_A t^3{dt\over y} &=& {3 \over 20}g_2 g_1 + {g_3\over 10},
  \quad\quad  {1\over g_0}\oint_B t^3{dt\over y} = \tau [{3 \over 20}g_2 g_1 + {g_3\over 
10}],
 \label{A3}
\end{eqnarray}
Here $g_1$ is also given by the Eisenstein series:
\begin{eqnarray}
g_1 = {1\over 3}({\pi\over g_0})^2[1 - 24\sum_{n=1}^{\infty}{nq^n\over 1-q^n}] \equiv 
-{1\over 3}({\pi\over g_0})^2 E_2(\tau). \label{C}
\end{eqnarray}
(\ref{A2}) and (\ref{A3}) can be easily shown from
$$
 \oint d(y) = 0, \quad\quad\quad \oint d(ty) = 0.
$$

\vspace{1cm}

\section{Calculations of $d\Omega_M$ and $\tau_{MN}$}
%\apsect{B}

\setcounter{equation}{0}

\noindent
We will obtain a formula for the $2$-point function $\tau_{MN}$, directly calculating the 
residue of (\ref{2.14}). First of all we expand $d\Omega_N$ given by 
  (\ref{2.7}) at $x ={1\over \xi^2} = \infty$. Let the expansion of ${dx\over y}$ to be
\begin{eqnarray}
{dx\over y} = -\sum^\infty_{M=1} R_M\xi^{2M-2}d\xi, \label{expand}
\end{eqnarray}
in which 
\begin{eqnarray}
&\quad& R_1 = 1, \quad\quad R_2 = {c\over 2}, \quad\quad R_3 = {5c^2\over 24} + {1\over 
8}g_2, \hspace{1cm} \nonumber \\
 &\quad& R_4 = {\ 35c^3\over 432} +{7c\over 48}g_2+{1\over 8}g_3,
 \label{R} \\ 
&\quad& R_5 = {35c^4\over 1152} + {7c^2\over 64}g_2 + {3c\over 16}g_3 + {3\over 
128}(g_2)^2, \nonumber \\
&\quad& R_6 ={77c^5\over 6912}+ {77c^3\over 1152}g_2+ {11c^2\over 64}g_3 + {11c\over 
256}(g_2)^2 + {3\over 64}g_3 g_2, \nonumber \\ 
&\quad& \hspace{1cm} \quad {\it etc}, \nonumber 
\end{eqnarray}
with $c$ defined by (\ref{defc}).  
Then the requirement (\ref{2.9}) gives the relation 
\begin{eqnarray}
   {}^t(\gamma_{N-1}\gamma_{N-2}\cdots\cdots \gamma_1) = -{\cal K}^{-1}{}^t(R_2\cdots\cdots 
R_N), \quad\quad N\ge 2 \label{A.4}
\end{eqnarray}
with  the matrix ${\cal K}$

\newfont{\bg}{cmr10 scaled \magstep4}
\newcommand{\bigzerol}{\smash{\hbox{\bg 0}}}
\newcommand{\bigzerou}{%
  \smash{\lower1.7ex\hbox{\bg 0}}}

\begin{eqnarray}
{\cal K} = 
\left(
\begin{array}{ccccc} 
R_1 &  &  &  &  \bigzerou   \\ 
R_2 & R_1 &  &  &   \\  
 \vdots & \vdots   &    & \ddots  &   \\      
R_{N-1} & R_{N-2} & \cdot  & \cdot & R_1  
\end{array}
\right).    \nonumber
\end{eqnarray}
By (\ref{2.7}) and (\ref{A.4}), (\ref{2.14}) is calculated as 
\begin{eqnarray}
{\scriptstyle{ 2M-1} \over\scriptstyle{ 2\pi i}}\tau_{NM} &=& 
-{\rm res}_{\xi=0}[\xi^{-2M+1}d\Omega_N] \nonumber \\
&=& R_{M+N} + \gamma_0 R_M   \nonumber \\
  &\quad& \quad -(R_{M+N-1}\cdots\cdots R_{M+1}){\cal K}^{-1}
  {}^t(R_2\cdots\cdots R_N),  \label{A.2}
\end{eqnarray}
for $N\ge 2$ and $ M\ge 1$. Here $\gamma_0$ is the constant in $d\Omega_N$ which is 
determined by (\ref{2.8}). For $N= 1$ and $ M\ge 1$ we have
\begin{eqnarray}
{\scriptstyle{ 2M-1} \over\scriptstyle{ 2\pi i}}\tau_{1M} &=&
-{\rm res}_{\xi=0}[\xi^{-2M+1}d\Omega_1] \nonumber \\
&=& R_{M+1} - (g_1 +{c\over 3})R_M,  \label{tau11}
\end{eqnarray}
in which $g_1$ was given by (\ref{C}). 
From $\tau_{1N} = \tau_{N1}$ we find alternatively $\gamma_0$ in (\ref{A.2}) to be 
\begin{eqnarray}
\gamma_0 &=& -{2N-2\over 2N-1}R_{N+1} - {1\over 2N-1}(g_1+{c\over 3})R_N   \nonumber  \\
  &\quad& \hspace{1cm}
       + (R_N\cdots\cdots R_2){\cal K}^{-1}{}^t(R_2\cdots\cdots R_N).
         \label{gamma0}, \quad\quad N=2,3,\cdots. \label{gamma00}
\end{eqnarray}
 For $\tau_{0N}$ with $ M\ge 1$ we have
\begin{eqnarray}
{\scriptstyle{ 2M-1} \over\scriptstyle{ 2\pi i}}\tau_{0M} \equiv  
{\scriptstyle{ 2M-1} \over\scriptstyle{ 2\pi i}}\tau_{M0} =
  { R_M \over g_0}, \label{A.3}
\end{eqnarray}
from (\ref{2.12}) and (\ref{2.13}). Evaluating the formulae (\ref{A.2}), (\ref{tau11}) and 
(\ref{A.3}) for simple cases 
 yields  $\tau_{AB}(=\tau_{BA})$ with $A,B= 0,1,2,\cdots$, as
\begin{eqnarray}
\tau_{01} &=& {2\pi i \over g_0}, \quad\quad 
\tau_{02} = {2\pi i \over g_0}{c\over 6}, \quad\quad 
\tau_{03} = {2\pi i \over g_0}[{c^2\over 24} + {g_2 \over 40}], \nonumber \\
\tau_{11} &=&  2\pi i [{c\over 6} - g_1 ],  \nonumber  \\
\tau_{12} &=&  2\pi i[{c^2 \over 72} -{c\over 6}g_1 + {g_2 \over 24}],    \nonumber \\
\tau_{13} &=& 2\pi i[{13c^3 \over 432}+{c^2\over 24}g_1+{3c \over 80}g_2 +{g_3 \over 40} 
+{g_1 g_2 \over 40}],   \label{2point} \\
\tau_{22} &=& \pi i[{ c^3\over 324}  - {c^2 \over 18}g_1 + {c\over 36} g_2 + {1\over 12}g_3 
],   \nonumber \\
\tau_{23} &=&  \pi i[{c^4\over 1728}-{c^3\over 72}g_1 + {c^2\over 96}g_2
 +{c\over 20}(g_3 -{g_1g_2\over 6}) + {(g_2)^2\over 192}],   \nonumber \\
\tau_{33} &=&  \pi i[ {c^5\over 8640} - {c^4\over 288}g_1 + {c^3\over 288}g_2 + {c^2\over 
240}(-g_2 g_1 + 6g_3)  \nonumber \\
&\quad& 
\hspace{1cm} + {c\over 192}(g_2)^2 +
{3\over 400}g_3 g_2  - {1\over 800}(g_2)^2 g_1 ].   \nonumber 
\end{eqnarray}

\vspace{1cm}

\section{Consistency check}
%\apsect{C}

\setcounter{equation}{0}

\noindent
  We shall alternatively  check that  (\ref{aa})  and (\ref{diftau}) follow from from 
(\ref{con}), although they are  proved by (\ref{kk}) together with (\ref{formula}) and the 
calculation (\ref{cal}) respectively. It gives an independent consistency check of our 
formalism for the Whitham hierarchy.

\subsection{Check of (7.12) }

%(\ref{aa})

To this end we have to explicitly evaluate $s_M(\tau)$, given by (\ref{speed}).  
 Let us put it in the form 
\begin{eqnarray}
s_M(\tau) = g_0 {\Gamma_M \over \Gamma_0}, \quad\quad M=1,2,\cdots,
 \label{sm}
\end{eqnarray}
with
\begin{eqnarray}
\Gamma_M =  u'Q_M(u)(v-w) + v'Q_M(v)(w-u) + w'Q(w)_M(u-v). \label{Gamma}
\end{eqnarray}
 Writing the curve (\ref{2.1}) in the Weierstrass standard form (\ref{C.1}) we have  
\begin{eqnarray}
u(\tau) = e_1 + {c\over 3}, \quad \quad v(\tau) = e_2 + {c\over 3}, \quad\quad
 w(\tau) = e_3 + {c\over 3}.  \label{stan}
\end{eqnarray}
Then  (\ref{Gamma}) reads
\begin{eqnarray}
\Gamma_M = (e_1' + {c'\over 3})Q_M(e_1 + {c\over 3})(e_2 -e_3) + cyclic.
 \label{C*}
\end{eqnarray}
First of all we replace $e_i'(={\partial e_i\over \partial \tau})$ in (\ref{C*})  by 
\begin{eqnarray}
{1\over 2\pi i}e_i' = -{1\over 2}({g_0\over \pi})^2( e_i^2 +g_1 e_i  -{g_2\over 6} )  
-{1\over \pi i}{g_0'\over g_0}e_i, \label{IM}
\end{eqnarray}
which is a variant  of the formula due to Itoyama and Morosov\cite{IM}
\begin{eqnarray}
2{\partial \over \partial \log q }\hat e_i = {1\over 6}\hat g_2(\tau) -\hat g_1(\tau)\hat 
e_i  - \hat e_i^2, \quad\quad\quad q = e^{2\pi i\tau}, \label{IMM}
\end{eqnarray}
with
\begin{eqnarray}
\hat e_i &=& ({g_0\over \pi})^2 e_i, \nonumber \\
\hat g_1(\tau) &=& -{1\over 3}E_2(\tau) = ({g_0\over \pi})^2 g_1, \nonumber \\
\hat g_2(\tau) &=& {2\over 3}[\theta_2(\tau)^8 + \theta_3(\tau)^8 + \theta_0(\tau)^8 ]  
\nonumber \\
 &=& {4\over 3}E_4(\tau) = ({g_0\over \pi})^4 g_2. \nonumber
\end{eqnarray}
Here $g_1$ and $g_2$ are given by (\ref{C}) and (\ref{g2}).
Then we find
\begin{eqnarray}
{1\over 2\pi i}\Gamma_M &=& [ -{1\over 2}({g_0\over \pi})^2( e_1^2 +g_1 e_1  -{g_2\over 6} 
) -{1\over \pi i}{g_0'\over g_0}e_1 + {1\over 2\pi i}{c'\over 3} ]
 Q_M(e_1 + {c\over 3})(e_2 -e_3) \nonumber \\
 &\quad& \hspace{1cm}  + cyclic.   \label{Gammaa}
\end{eqnarray}
Next we replace  $Q_M(e_i + {c\over 3}), i = 1,2,3$, in the {\it r.h.s.} by those 
calculated in (\ref{dSS}), (\ref{dS2}) and (\ref{dS3}).  
The resultant $\Gamma_M$  is simplified  by  the formulae 
\begin{eqnarray}
{1\over 2}({g_0\over \pi})^2e_1^2(e_2-e_3) + cyclic &=& -{\Gamma_0\over 2\pi i}, \label{f1} 
\\
{1\over 2}({g_0\over \pi})^2e_1^3(e_2-e_3) + cyclic &=& 0, \label{f2}\\
{1\over 2}({g_0\over \pi})^2e_1^4(e_2-e_3)+ cyclic &=&
 -{\Gamma_0\over 2\pi i}{g_2\over 4}, \label{f3}  \\
{1\over 2}({g_0\over \pi})^2e_1^5(e_2-e_3)+ cyclic &=&
 -{\Gamma_0\over 2\pi i}{g_3\over 4}, \label{f4} \\
 {1\over 2}({g_0\over \pi})^2 e_1^6(e_2-e_3)+ cyclic &=&  
 -{\Gamma_0\over 2\pi i}({1 \over 16}(g_2)^2 +{1\over 2}g_1g_3),   \label{f5}  \\
  \cdots\cdots & &  \nonumber
\end{eqnarray}
Finally we eliminate  $c'(={\partial c\over \partial \tau})$ 
 from $\Gamma_M$ for the cases with $M_0=1, M_0 =2$ and $M_0=3$ by means of 
the constraints (\ref{con1})$\sim$(\ref{con3}) respectively. To this end we need to 
differentiate the constraints by $\tau$  using the  formulae 
\begin{eqnarray}
{\partial \over \partial \tau} E_2(\tau) &=& {i\pi \over 6}( E_2(\tau)^2 - E_4(\tau)), 
\label{s1}  \\
{\p \over \p \tau}E_4(\tau) &=& {2\pi i\over 3}( E_2(\tau)E_4 -E_6(\tau) ), \label{s2}   \\
{\partial \over \partial \tau} E_6(\tau) &=& -i\pi ( E_4(\tau)^2 - E_2(\tau)E_6 (\tau)), 
\label{s3}  \\
 & &  \cdots\cdots.  \nonumber 
\end{eqnarray}
We have checked that the characteristic speeds (\ref{sm}) with $\Gamma_M$ thus evaluated 
indeed satisfy (\ref{aa}) for the respective case of $M_0=1, M_0 =2$ and $M_0=3$.

\vspace{0.5cm}

The formulae 
 (\ref{f1})$\sim$(\ref{f5}), (\ref{s1})$\sim$(\ref{s3}) can be shown  by means of 
(\ref{IM}). First of all we show (\ref{f1})$\sim$(\ref{f5}). 
Integrating (\ref{difholo}) along a $A$-cycle in the Weierstrass standard form we obtain 
$$
\oint_A{1\over t-e_1}{dt\over y} = {g_0\over \Gamma_0}(e_2'-e_3') + 2{e_2-e_3\over 
\Gamma_0}g_0'. 
$$
By means of (\ref{IM}) it becomes 
\begin{eqnarray}
{1\over 2\pi i}\oint_A{1\over t-e_1}{dt\over y} = {1\over 2}({g_0\over \pi})^2{g_0\over 
\Gamma_0}(e_2-e_3)(e_1-g_1). \label{int}
\end{eqnarray}
The similar formulae are found by cyclic rotation of $e_1$, $e_2$ and $e_3$. 
Now we calculate $d({t^n\over y})$ for $n=1,2,3,4,5$ as
\begin{eqnarray}
&\quad& d({t\over y}) = [-{1\over 2}-{1\over 2}\sum_i {e_i\over t-e_i}]{dt\over y}, 
\hspace{2cm} \nonumber \\
&\quad&d({t^2\over y}) = [{t\over 2}-{1\over 2}\sum_i {e_i^2\over t-e_i}]{dt\over y}. 
\hspace{2cm} \nonumber \\
&\quad& \hspace{1cm} \cdots\cdots\cdots\cdots    \nonumber
\end{eqnarray}
Integrate them along a $A$-cycle and use (\ref{int}). With (\ref{g2})$\sim$(\ref{C}) we 
then obtain the formulae (\ref{f1})$\sim$(\ref{f5}).

\vspace{0.5cm}

Next  we go to the proof of (\ref{s1})$\sim$(\ref{s3}). Either of them is proved similarly. 
We will sketch a proof of (\ref{s2}). Differentiate the first equation of (\ref{A2}) by 
$\tau$:
\begin{eqnarray}
({g_2 \over 12})' = -{g_0'\over g_0}{g_2 \over 12} +
 {1\over 2 g_0}\sum_ie_i'\oint_A [t+e_i + {e_i^2\over t-e_i}]{dt\over y}.
  \label{difg}
\end{eqnarray}
We calculate the second piece of the {\it r.h.s.} by the following three steps:
 At first perform the integration by (\ref{A1})$\sim$(\ref{A3}) and (\ref{int}). Next 
replace $e_i'$ by the formula (\ref{IM}). Finally sum over $i$ by 
(\ref{f1})$\sim$(\ref{f4}). Then we find 
\begin{eqnarray}
 {1\over 2 g_0}\sum_ie_i'\oint_A [t+e_i + {e_i^2\over t-e_i}]{dt\over y}\ = \
 -{\pi i \over 2}({g_0\over \pi})^2[{g_3\over 2} + {g_1g_2\over 3}] -
{g_0'\over g_0}{g_2\over 4}.  \label{2nd}
\end{eqnarray}
On the other hand the {\it l.h.s.} of (\ref{difg}) is calculated as 
\begin{eqnarray}
({g_2 \over 12})' = -{g_0'\over g_0}{g_2\over 3} + {1\over 9}({\pi\over g_0})^4E_4(\tau)'
 \label{lhs},
\end{eqnarray}
by (\ref{g2}). Putting  (\ref{2nd}) and (\ref{lhs}) into (\ref{difg}) and using the 
Eisenstein series, we obtain (\ref{s2}).

\subsection{Check of (7.32) }

%(\ref{diftau})

We check (\ref{diftau}) for case ii), 
{\it i.e.},
\begin{eqnarray}
\Lambda_2^2{\p \over \p \tau}\tilde\tau_{22} = [{\pi\over 72}(E_2(\tau)^2 - E_4(\tau))]^2.  
\label{D.1}
\end{eqnarray}
 For other cases it can be done similarly.  
In the first place we compute  the residue in $\tilde\tau_{22}$ given by (\ref{22}):
\begin{eqnarray}
 &2\pi& i \ {\rm res}_{\xi=0}[\xi^{-4}d\xi \Delta_2(x,\tau)]  \nonumber \\
 &=&  \pi i[-{1\over \Gamma_0}\{Q_2(u)u'(v'w-w'v) + cyclic\} 
        +{c\over 2\Gamma_0}\{Q_2(u)u'(v'-w') + cyclic\} ], \nonumber \\
                      \label{D.2}
\end{eqnarray}
using (\ref{exact}) and (\ref{expand}). 
By (\ref{stan}) both cyclic sums can be written as 
\begin{eqnarray}
Q_2&(u)&u'(v'w-w'v) + cyclic   \nonumber \\
&=& Q_2(e_1 + {c\over 3})(e'_1 + {c'\over 3})  
[ (e_2'e_3-e_3'e_2) +{c\over 3}(e_2'-e_3') -{c'\over 3}(e_2-e_3) ], \nonumber 
\\
Q_2&(u)&u'(v'-w') + cyclic   \nonumber \\
&=& Q_2(e_1 + {c\over 3})(e'_1 + {c'\over 3}) (e_2'-e_3'). \nonumber
\end{eqnarray}
With recourse to (\ref{IM}) and (\ref{f1})$\sim$(\ref{f5})  they are evaluated in the same 
way as $\Gamma_M$, (\ref{C*}). After tedious calculations we  find that 
\begin{eqnarray}
 2\pi i \ &{\rm res}_{\xi=0}&[\xi^{-4}d\xi \Delta_2(x,\tau)]  \nonumber \\
 &=&  (\pi i)^2 ({ g_0\over \pi})^2 [{g_2^2 \over 72} + {g_1 g_3 \over 4}+
 + c({g_1 g_2 \over 18} + {g_3 \over 12}) + c^2({g_2 \over 432} -{(g_1)^2 \over 36}) ]    
\nonumber \\
&\quad& + 2 \pi i {g_0' \over g_0}[{g_3\over 4} + {c\over 18}g_2 -{c^2\over 18}g_1 ]   + 
\pi i c' [ -{g_2\over 36} +{c\over 9}g_1 - {c^2\over 108} ] \nonumber\\&\quad& - ({\pi 
\over g_0})^2 [{1\over 3}(c'-c{g_0'\over g_0})]^2. \label{res}
\end{eqnarray}
On the other hand we calculate ${\p \tau_{22}\over\p \tau}$ by differentiating directly the 
formula given by (\ref{2point}). Then add  the result to  (\ref{res}). It exactly cancels 
the first three terms of (\ref{res}). We use the constraint (\ref{M2}) and (\ref{s1}) in 
the last term to obtain (\ref{D.1}).

\end{document}